\documentclass[fleqn,usenatbib]{mnras} 

\usepackage{newtxtext,newtxmath}

\usepackage[T1]{fontenc}
\usepackage{ae,aecompl}

\usepackage{graphicx}	
\usepackage{amsmath}	

\usepackage{amssymb}	
\usepackage{multicol}
\usepackage{bm}
\usepackage{booktabs}
\usepackage{cleveref}
\usepackage{natbib}
\usepackage{threeparttable}

\newcommand{\secpoint}{\mbox{$''\mskip-7.6mu.\,$}}
\newcommand{\angstrom}{\mbox{\normalfont\AA}}

\newcommand{\muv}{$\rm M_{\rm UV}$}

\newcommand{\jwst}{{\it JWST}}

\newcommand{\rxc}{RXCJ2248-ID}
\newcommand{\gnz}{GN-z11}

\defcitealias{Bian2018}{B18}
\defcitealias{PP04}{PP04}

\title[{High-ionization emission lines at $z>6$}]{Metal-poor star formation at $z>6$ with {\it JWST}: new  insight into hard radiation fields and nitrogen enrichment on 20 pc scales    }

\author[M. W. Topping]{Michael W. Topping$^{1}$\thanks{E-mail: michaeltopping@arizona.edu},
Daniel P. Stark$^{1}$,
Peter Senchyna$^{2}$,
Adele Plat$^{1}$,
Adi Zitrin$^{3}$,\newauthor
Ryan Endsley$^{4}$,
St\'ephane Charlot$^{5}$,
Lukas J. Furtak$^{3}$,
Michael V. Maseda$^{6}$,
Renske Smit$^{7}$,\newauthor
Ramesh Mainali$^{8}$,
Jacopo Chevallard$^{9}$,
Stephen Molyneux$^{7,10}$,
Jane R. Rigby$^{11}$
\\
$^{1}$Department of Astronomy / Steward Observatory, University of Arizona, 933 N Cherry Ave, Tucson, AZ 85721\\
$^{2}$The Observatories of the Carnegie Institution for Science, 813 Santa Barbara Street, Pasadena, CA 91101, USA\\
$^{3}$Physics Department, Ben-Gurion University of the Negev, P.O. Box 653, Be’er-Sheva 84105, Israel\\
$^{4}$Department of Astronomy, University of Texas, Austin, TX 78712, USA\\
$^{5}$Sorbonne Universit\'e, CNRS, UMR 7095, Institut d’Astrophysique de Paris, 98 bis bd Arago, 75014 Paris, France\\
$^{6}$Department of Astronomy, University of Wisconsin-Madison, 475 N. Charter St., Madison, WI 53706, USA\\
$^{7}$Astrophysics Research Institute, Liverpool John Moores University, 146 Brownlow Hill, Liverpool L3 5RF, UK\\
$^{8}$Observational Cosmology Lab, Code 665, NASA Goddard Space Flight Center, 8800 Greenbelt Rd., Greenbelt, MD 20771, USA\\
$^{9}$Department of Physics, University of Oxford, Denys Wilkinson Building, Keble Road, Oxford OX1 3RH, UK\\
$^{10}$European Southern Observatory, Karl-Schwarzschild-str. 2, 85748 Garching, Germany\\
$^{11}$Astrophysics Science Division, Code 660, NASA Goddard Space
Flight Center, 8800 Greenbelt Rd., Greenbelt, MD, 20771, USA.
}

\begin{document}
\label{firstpage}
\pagerange{\pageref{firstpage}--XX}
\maketitle
\begin{abstract}
Nearly a decade ago, we began to see indications that reionization-era galaxies power hard radiation fields rarely seen at lower redshift. Most striking were detections of nebular CIV emission in what appeared to be typical low mass galaxies, requiring an ample supply of 48 eV photons to triply ionize carbon. We have obtained deep JWST/NIRSpec R=1000 spectroscopy of the two z>6 CIV-emitting galaxies known prior to JWST. Here we present a rest-UV to optical spectrum of one of these two systems, the multiply-imaged $z=6.1$ lensed galaxy RXCJ2248-ID. NIRCam imaging reveals two compact ($<22pc$) clumps separated by 220pc, with one comprising a dense concentration of massive stars ($>10,400M_{\odot}$/yr/kpc$^2$) formed in a recent burst. We stack spectra of 3 images of the galaxy (J=24.8--25.9), yielding a very deep spectrum providing a high S/N template of strong emission line sources at $z>6$. The spectrum reveals narrow high ionization lines (HeII, CIV, NIV]) with line ratios consistent with powering by massive stars. The rest-optical spectrum is dominated by very strong emission lines ([OIII] EW=2800\AA), albeit with weak emission from low-ionization transitions ([OIII]/[OII]=184). The electron density is found to be very high ($6.4-31.0\times10^4$cm$^{-3}$) based on three UV transitions. The ionized gas is metal poor ($12+\log(\rm O/H)=7.43^{+0.17}_{-0.09}$), yet highly enriched in nitrogen ($\log(\rm N/O)=-0.39^{+0.11}_{-0.10}$). The spectrum appears broadly similar to that of GNz11 at $z=10.6$, without showing the same AGN signatures. We suggest that the hard radiation field and rapid nitrogen enrichment may be a short-lived phase that many $z>6$ galaxies go through as they undergo strong bursts of star formation. We comment on the potential link of such spectra to globular cluster formation.

\end{abstract}


%
%
%
%
\section{Introduction} 
\label{sec:intro}

The rest-frame ultraviolet (rest-UV) provides powerful diagnostics of the ionizing sources and gas conditions in galaxies. The massive O and B stars present in galaxies 
will contribute significantly to the continuum, while also producing several P-Cygni wind lines and numerous photospheric absorption lines that are sensitive to the metallicity and age of the stellar population \citep{Shapley2003, Leitherer2011, Crowther2016, Steidel2016, Rigby2018}. Resonant metal transitions throughout the UV probe interstellar and circumgalactic gas in absorption, constraining  
the covering fraction and kinematics of the outflowing gas that regulates the escape of Lyman Continuum and Ly$\alpha$ emission. The presence of emission features from numerous 
high ionization species (NV, NIV, CIV, He II)  constrain the shape of the radiation field, 
providing a signpost of hard spectra associated with active galactic nuclei (AGNs) or very hot stellar populations. 
Finally the metal emission lines in the rest-UV constrain both the nitrogen and carbon abundance (relative to oxygen) and the electron density of the higher ionization gas.

The first glimpse of the rest-UV emission spectra of $z\gtrsim 6$ galaxies came prior to {\it JWST}, revealing 
hints of significant evolution from lower redshifts. Deep near-infrared spectra of bright ($H\simeq 25$) galaxies with ground-based telescopes presented detections of CIII]$\lambda\lambda$1907,1909 emission \citep{Stark2015a, Stark2017, Laporte2017, Hutchison2019, Topping2021} with rest-frame equivalent widths (EWs) that were largely unparalleled among samples at lower redshifts \citep{Erb2010, Stark2014, deBarros2016, Amorin2017, Maseda2017, Berg2018, Feltre2020, Du2020, Tang2021}.  Yet more striking were two detections of  
CIV$\lambda\lambda$1548,1550 emission lines in what appeared to be typical star forming galaxies at $z\gtrsim 6$. These observations indicated a radiation field  capable of triply ionizing carbon (ionization potential = 48 eV) in the vicinity of the ionizing sources. The first CIV detection was discovered 
in a bright (H=25.8) gravitationally lensed galaxy at $z=7.045$ galaxy in Abell 1703 \citep{Stark2015b}, previously shown to have Ly$\alpha$ emission in \citet{Schenker2012}.  The second detection came shortly 
after in the $z=6.11$ gravitationally lensed galaxy, RXCJ2248-ID \citep{Mainali2017, Schmidt2017}, also bright (multiple images ranging between J=24.8 and 25.9) and previously known to have Ly$\alpha$ emission \citep{Balestra2013}.  
The fact that these lines were discovered in two of the first galaxies with deep spectra sampling the CIV complex gave an early indication that intense radiation fields may be  ubiquitous in the reionization era.

The origin of the high ionization lines in the $z\gtrsim 6$ Universe remains unclear. At lower redshifts, CIV emission has long been 
associated with AGNs, and indeed some argued that the $z\gtrsim 6$ CIV detections may be best explained by an AGN power law spectrum in the EUV \citep{Laporte2017, Nakajima2018}. If true, these detections may have provided the first 
evidence that AGNs are fairly common in typical (UV-selected) galaxies at $z\gtrsim 6$, a conclusion now more widespread in the {\it JWST} era \citep{Greene2023, Larson2023, Furtak2023, Fujimoto2023b,Maiolino2023b,Scholtz2023}.  However others have argued that 
the origin of the high ionization emission may instead be a very low metallicity population of massive stars \citep{Stark2015b,Mainali2017}. This interpretation 
was driven by the rest-UV line ratios which indicate 
weaker He II emission than might be expected from narrow-line AGN photoionization \citep{Mainali2017}. 
Furthermore, this physical picture was bolstered by indications of hard radiation fields in  existing observations of nearby metal poor galaxies thirty years ago \citep[e.g.,][]{Garnett1991, Thuan2005}. However the majority of the local metal poor galaxies studied in the UV prior to 2015 lacked coverage 
extending down to CIV, prohibiting a direct comparison with the emerging $z\gtrsim 6$ emission line spectra.  

To overcome the deficiencies  in our knowledge of  low metallicity ($\lesssim 0.1$ Z$_\odot$) massive star populations, a series of {\it HST} UV surveys using the Cosmos Origins Spectrograph (COS) 
were undertaken over the past decade to characterize nearby ($\lesssim$ 100 Mpc) metal poor galaxies \citep{Senchyna2017, Berg2019b,Senchyna2019,Wofford2021}. The observations demonstrated that 
the UV emission lines seen at $z\gtrsim 6$ (i.e., CIII], CIV) do become stronger in lower metallicity galaxies with young stellar populations. 
Physically  such a trend may naturally arise from the harder radiation field in low metallicity massive stars and the stronger collisionally-excited lines at the higher electron temperatures in metal poor HII regions \citep{Senchyna2017, Senchyna2019, Berg2019, Berg2019b, Schaerer2022}. Spectroscopic surveys at $z\simeq 2-3$ have put forward a similar picture \citep{Tang2021, Saxena2022}. But while  high ionization lines have been located in $z\simeq 0-2$ metal poor star forming galaxies, the CIV EWs have largely been well below those seen at $z\gtrsim 6$. Only one recently studied system presents a CIV line detection with a rest-frame EW (43~\AA) reaching the level observed in reionization-era systems \citep{Izotov2024}.

Attention has recently shifted back to reionization-era spectroscopy following the launch of {\it JWST}.  
Most early spectroscopic work has focused on  the emission lines in the rest-frame optical at $z\gtrsim 6$. The  optical emission lines provide well-established diagnostics of the ionized gas and ionizing sources, with a series of ionization and temperature sensitive lines. 
The first results have revealed many $z\gtrsim 6$ galaxies have 
ionized gas that is fairly metal poor (0.05-0.30 Z$_\odot$) and 
under  extreme ionization conditions \citep{Sanders2023, Cameron2023, Tang2023, Mascia2023}. The spectra appear consistent with lower redshift scaling relations between rest-optical emission line EWs and ionization-sensitive ratios \citep[e.g.,][]{Tang2019,Sanders2020,Tang2023}.
The rest-frame EWs of [OIII] and H$\alpha$ at $z\gtrsim 6$ tend to be much larger than those in lower redshift samples \citep{Endsley2023b}, as expected for galaxies dominated by young stellar populations formed in a recent burst of star formation. The high ionization state of $z\gtrsim 6$ HII regions likely reflects conditions associated with powerful bursts which appear common in early galaxies.

Progress in the rest-UV with {\it JWST} has been slow to-date. 
Since the rest-UV emission lines are weaker than those in the rest-optical, the majority of $z\gtrsim 6$ galaxies targeted with {\it JWST} 
in Cycle 1 have been too faint ($H\simeq 27-30$) for useful constraints. Only spectra of the brightest continuum ($H\simeq 24-26$) sources can reach the CIII] and CIV  rest-frame EWs ($\simeq 10$-30~\AA) found in earlier ground-based investigations. Several bright $z\gtrsim 6$ galaxies with Cycle 1 spectra have indeed shown strong rest-UV emission lines \citep{Tang2023, Hsiao2023, Fujimoto2023b}. The poster child of these sources is GNz11 \citep{Oesch2016}. NIRSpec observations presented in \citet{Bunker2023} reveal a suite of 
powerful emission lines, including the high ionization rest-UV lines discussed above (i.e., CIV, CIII], OIII], NIV]). The nitrogen lines point to gas that is both metal poor and nitrogen-enhanced, an   abundance pattern that is extremely uncommon in star forming galaxies at lower redshifts. Several authors have 
pointed out that it is  similar to that of the second-generation stellar populations in globular clusters (GCs; e.g., \citealt{Bastian2018}), leading to the suggestion that GNz11 may be linked to an evolutionary phase associated with globular cluster formation \citep{Senchyna2023}.  The doublet flux ratios in GNz11 point to extremely large electron densities  ($\gtrsim 10^{5}$ cm$^{-3}$)  rarely seen in the HII regions of star forming galaxies at lower redshifts, leading to suggestions we may be seeing emission from  the broad line region (BLR) of an AGN \citep{Maiolino2023}. 

Whether spectra like GNz11 are common among the early galaxy population is not clear.  The high ionization lines, enhanced electron density, and peculiar nitrogen abundance pattern  may  correspond to an evolutionary phase that many $z\gtrsim 6$ galaxies go through, either shortly after a burst of star formation or during an AGN episode. 
A natural first step is testing whether the galaxies previously known 
to have high ionization UV lines at $z\gtrsim 6$ also show the enhanced 
electron density and peculiar abundance pattern seen in GNz11. In {\it JWST} Cycle 1, spectroscopic observations of both $z\gtrsim 6$ galaxies with CIV emission detected from the ground were obtained in GO 2478 (PI: Stark). As both CIV emitters are very bright continuum sources (10--100$\times$ greater than typical sources in deep fields), this dataset will provide valuable spectroscopic templates for understanding emission lines in reionization era galaxies. In this paper, we present initial results from this {\it JWST} program, focusing on observations of the $z=6.11$ gravitationally lensed CIV emitter in the cluster field 
RXC J2248.7-4431 \citep{Mainali2017,Schmidt2017}. We have obtained moderate resolution (R=1000) NIRSpec spectra spanning the rest-UV to optical for three images of the galaxy, with apparent magnitudes (J=24.8--25.9) enabling spectra with continuum S/N in excess of  that for GNz11.  Our goals for this first paper are twofold. We  aim to provide new insight into the ionizing sources (i.e., stellar populations, AGN, shocks) and gas properties (i.e., metallicity, ionization state, density) linked to high ionization line detections in reionization era. We also seek to understand more about what physical conditions and evolutionary stages produce spectra similar to GNz11.

This structure of this paper is as follows.
Section~\ref{sec:data} provides an outline of the observations and data analysis, and discusses our measurement methods.
We explore the basic physical properties implied by new {\it JWST} broadband flux measurements in 
Section~\ref{sec:stars}.
In Section~\ref{sec:results} we describe the emission and absorption lines detected in the NIRSpec observations, and in Section \ref{sec:derived} we use these measurements to infer gas-phase properties of the CIV emitter. 
Section~\ref{sec:disc} discusses implications for our findings.
Finally, we provide a summary and brief conclusions in Section~\ref{sec:summary}.
Throughout this paper we assume a cosmology with $\Omega_m = 0.3$, $\Omega_{\Lambda}=0.7$, $H_0=70 \textrm{km s}^{-1}\ \textrm{Mpc}^{-1}$, adopt solar abundances from \citet[][i.e., $Z_{\odot}=0.014$, $12+\log(\rm O/H)_{\odot}=8.69$]{Asplund2009}, present magnitudes using the AB system \citep{Oke1984}, and provide equivalent widths in the rest frame. Line wavelengths are defined using the NIST Atomic Spectra Database.

\begin{figure}
    \centering
     \includegraphics[width=1.0\linewidth]{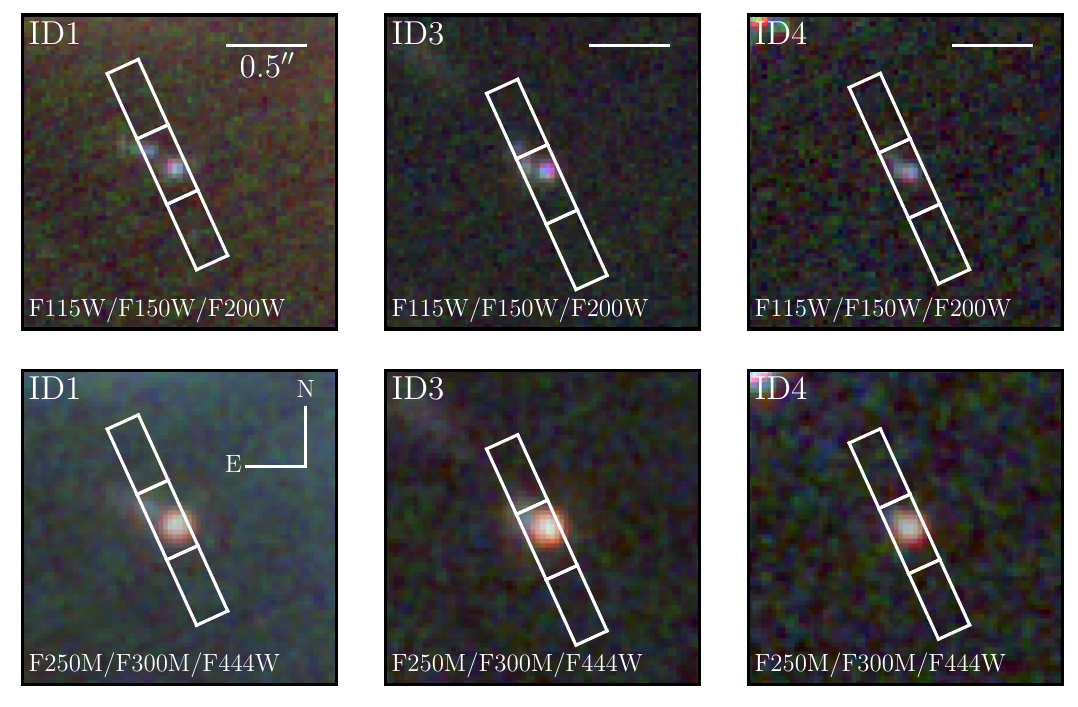}
     \caption{NIRCam data of the three observed lensed images of RXCJ2248-ID. The top row displays the rest-UV colors (F200W/F150W/F115W for red/green/blue), while the bottom row shows the rest-optical colors (F250M/F300M/F444W for red/green/blue).} 
     \label{fig:images}
\end{figure}

%
%
%
%
\section{Data and Measurements}
\label{sec:two}

In this section we present NIRSpec observations of the gravitationally lensed  CIV emitting galaxy at $z=6.11$ behind the RXCJ2248-4431 cluster.
Throughout this paper, we will refer to this system as RXCJ2248-ID, while the five individual lensed images are presented as RXCJ2248-ID1, -ID2, -ID3, -ID4, and -ID5, following the terminology of earlier papers. 
RXCJ2248-ID was first identified as a multiply-lensed, bright ($\rm J_{\rm AB}=24.8-25.9$) $z\sim6$ candidate  \citep{Monna2014, Boone2013} from the CLASH survey \citep{Postman2012}.
Its redshift was subsequently confirmed to be $z_{\rm spec}=6.11$ based on detection of Ly$\alpha$ \citep{Balestra2013}.
Subsequent spectroscopy of RXCJ2248-ID identified detections of CIV$\lambda$1550 and OIII]$\lambda$1666 
\citep{Mainali2017,Schmidt2017}, revealing the presence of hard 
ionizing sources in this  system. 
The NIRSpec observations presented in this paper are aimed at providing a deeper view of the emission lines in the rest-frame UV to optical. 
In this section we provide an overview of the program from which our data derive (Section~\ref{sec:observations}), along with the methods used to measure emission and absorption lines from the spectra (Section~\ref{sec:measurements}).
\label{sec:data}

\subsection{Observations and Reduction}
\label{sec:observations}
Spectroscopic data were obtained using \jwst{}/NIRSpec in Multi-Object Spectroscopy (MOS) mode targeting the RXCJ2248-4431 lensing field \citep[RA: 342.17972, Dec: -44.5330,][]{Guzzo2009}, as part of Cycle 1 program ID 2478 (PI: Stark).
Initial observations in the RXCJ2248-4431 field in November 2022 were affected by an electrical short in the microshutter array, which is a known issue, and were subsequently re-executed and completed in May 2023. 
For this program, we obtained spectra of RXCJ2248-ID1 (J=24.8),  RXCJ2248-ID3 (J=25.0), and RXCJ2248-ID4 (J=25.9).
We display NIRCam images along with the positions of the NIRSpec shutters for each targeted image in Figure~\ref{fig:images}. The NIRCam images of RXCJ2248-4431 were obtained as part of Cycle 1 Program 1840 (PI: Alvarez-Marquez), and comprise short-wavelength (SW) imaging in F115W, F150W, and F200W, as well as long-wavelength (LW) imaging in F250M, F300M, and F444W.

The NIRSpec observations reported in this paper used the G140M/F100LP, G235M/F170LP, and G395M/F290LP gratings and filters and utilized the NRSIRS2RAPID readout mode.
Total exposure time in these three setups was 6215 sec., 1576 sec., and 1576 sec., respectively.
Each slit placed upon the targets was composed of three microshutter slitlets, and we obtained data using the standard three-nod pattern, with one exposure at each nod position.

Final spectra were derived using a data reduction pipeline composed of standard STScI tools\footnote{\url{https://github.com/spacetelescope/jwst}} \citep{Bushouse2023} in addition to custom routines described below.
The first reduction steps were applied to the raw uncalibrated 
(\texttt{*\_uncal.fits}) 
frames which included flagging of cosmic rays and ramp fitting in addition to subtraction of significant events such as `snowballs' and `showers'.
The resulting images were corrected for $1/f$ noise using \texttt{nsclean} \citep{Rauscher2023}, and the 2D spectrum of each individual object was cut out from each image.
Following this, we applied a flat-field correction, applied a wavelength solution and absolute photometric calibration using the updated Calibration Reference Data System (CRDS) context.
Each nodded exposure of the targets were then background subtracted and combined including the rejection of pixels that have been flagged in previous reduction stages.
Final 2D spectra were then interpolated onto a common wavelength grid.
We fit the spatial profile of each 2D spectrum, and used the fitting information to perform an optimal extraction \citep{Horne1986} yielding the final 1D science and error spectra. Finally, we verified that the error spectrum was consistent with the rms scatter measured in a blank region of the science spectrum.

We correct the emission from each lensed image for the effects of magnification.
We revise the lens model by \citep{Zitrin2015} with an updated version of their parametric method, which is no longer coupled to a fixed grid resolution and thus capable of high resolution results.
As constraints we also use a more updated set of 14 multiply imaged galaxies and knots from the compilation by \citet{Caminha2016}, the majority of which have measured spectroscopic redshifts. 
The model consists of two main components: cluster galaxies and dark matter halos. Cluster galaxies are modeled each as a double Pseudo Isothermal Ellipsoid (dPIE; \citealt{Eliasdottir2007}), scaled by its luminosity following common scaling relations \citep{Jullo2007}. 
We model the dark matter halos each as Pseudo Isothermal Elliptical Mass Distribution (PIEMD; e.g., \citealt{Keeton2001}), where two halos are used, one centered on the brightest cluster galaxy (BCG), and on a second bright cluster galaxy north-east of the main one (RA:342.21621, DEC:-44.518370).  
The mass of the BCG is modeled independently of the scaling relation, allowing some more freedom in the core, and the redshifts of the photometric-redshift systems are left as free parameters in the minimization as well.
The minimization is performed via a long MCMC and the final model has an image reproduction rms of 0\secpoint4.

Image ID1 has the largest magnification ($\mu=7.9$), and appears as two distinct components separated by $\sim0\secpoint24$ in the image plane, corresponding to a physical distance of $\sim220~\rm pc$ ($0\secpoint04$) in the source-plane reconstruction.
Both ID1 components are unresolved ($r_{\rm eff}<0\secpoint 03$) and have equal lensing magnifications which allow their sizes to each be constrained to just $r_{\rm eff}\lesssim 22$~pc.
The southwest component (in ID1; see Figure~\ref{fig:images}) comprises $70\%$ of the emission of the system at $1.15\rm \mu m$ (i.e., in NIRCam/F115W), and forms the primary target of our spectroscopy in each observed image.

We stacked the spectra of the observed lensed images.
The spectrum of each image was interpolated onto a common wavelength grid, and de-magnified following the lens model.
Following this step, we confirmed that each of the three spectra had a consistent normalization, implying that they are on the same absolute flux scale.
The resulting spectra were subsequently combined at each wavelength using a weighted average, and excluding wavelengths that fell between the NIRSpec detectors.
This process yielded continuous wavelength coverage from 9790~\angstrom{} -- 52000~\angstrom{} in the observed frame.
Our final combined spectrum of RXCJ2248-ID presents a continuum S/N of 6 per resolution element (resel), where one resel corresponds to a width of 
$300~\rm km/s$. 
The $3\sigma$ limiting line flux that we obtain is $1.2\times10^{-19}~\rm erg/s/cm^2$ at 1$\rm \mu m$.
At $5\rm \mu m$ we detect the continuum at a S/N of 2 per resel, and we obtain a $3\sigma$ limiting line flux of $1.5\times10^{-19}~\rm erg/s/cm^2$.  At these sensitivities, we 
can reach rest-frame EWs of 1.2~\AA\ (rest-UV) and 18~\AA\ (rest-optical) at the $3\sigma$ level.

\begin{figure}
    \centering
     \includegraphics[width=1.0\linewidth]{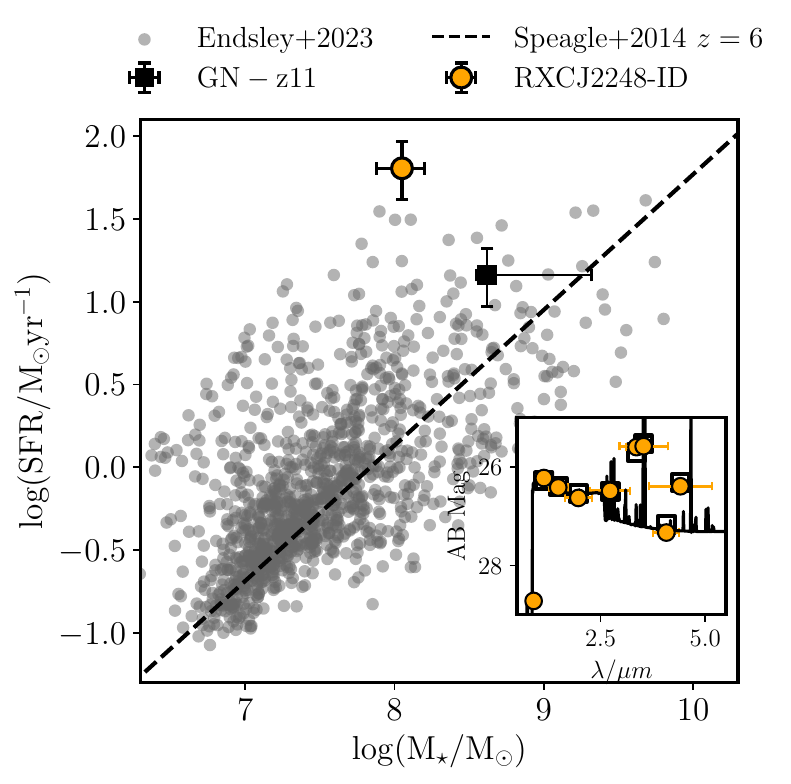}
     \caption{Star-formation rate versus stellar mass for RXCJ2248-ID (orange point) inferred from the BEAGLE SED fitting as described in Section~\ref{sec:stars}. We compare to the values inferred for GN-z11 \citep[black square;][]{Bunker2023}, in addition to a population of star-forming galaxies at $z\sim6-9$ identified by \citet{Endsley2023} (grey circles). For reference, we display the `main sequence' derived by \citet{Speagle2014} calculated at $z=6$. Our primary targets lie at significantly elevated SFR relative to the main sequence and the population of early star-forming galaxies at fixed stellar mass. The inset panel shows the photometric points synthesized from the spectrum and best-fit SED model as described in Section~\ref{sec:stars}.} 
     \label{fig:mainsequence}
\end{figure}

\begin{figure*}
    \centering
     \includegraphics[width=1.0\linewidth]{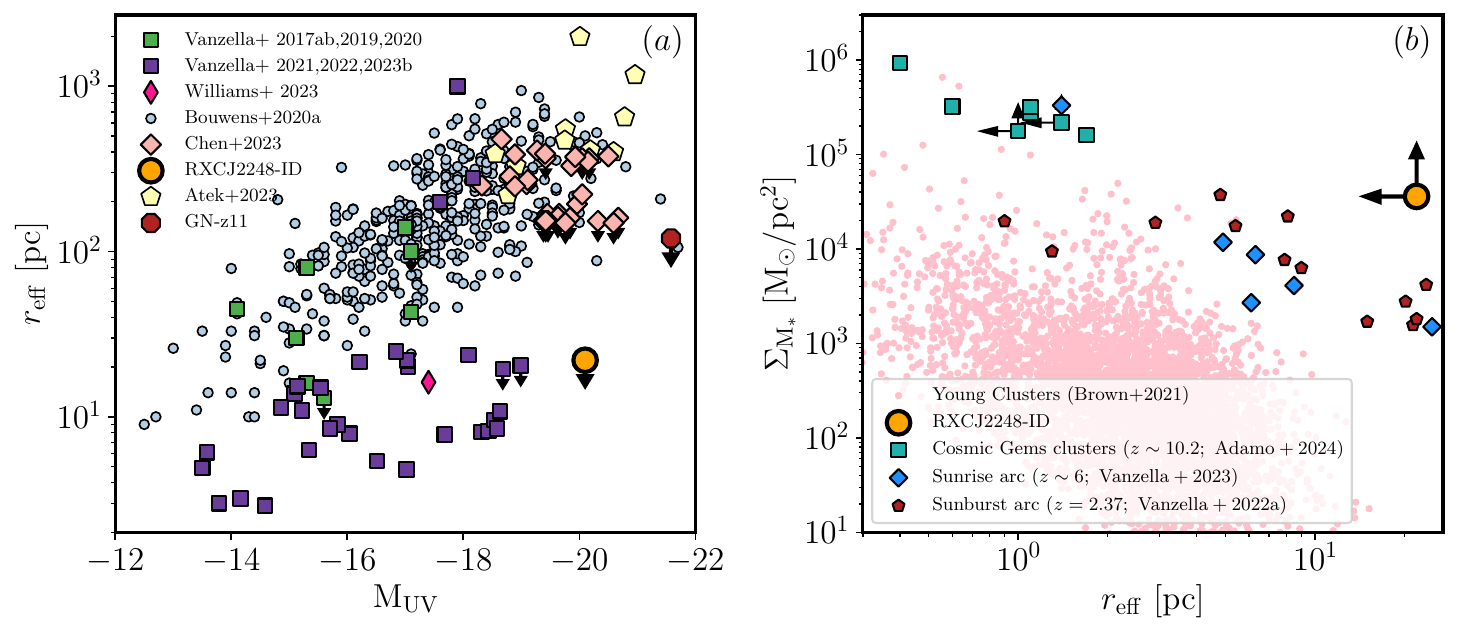}
     \caption{Panel (a): Effective radii ($r_{\rm eff}$) versus \muv{} measured for RXCJ2248-ID (orange circle) compared to measurements from the literature. The light blue circles are derived from gravitationally lensed galaxies in the {\it Hubble} Frontier Fields \citet{Bouwens2021}, and shows an increase in galaxy radius toward larger UV luminosity. The pink diamonds represent individual star-forming complexes at $z\sim6-8$ from \citet{Chen2023}, and a sample of lensed galaxies at $z\sim9-16$ are indicated by yellow pentagons. RXCJ2248-ID is significantly more compact than these other high-redshift systems with measured sizes at the same UV luminosity. Measurements of GN-z11 \citep[red hexagon][]{Tacchella2023}, RXJ2129-z95 \citep{Williams2023}, and lensed protoglobular clusters \citep[blue squares;][]{Vanzella2022,Vanzella2023} show a similar offset toward more compact sizes to that of RXCJ2248-ID, indicating they may be in a similar mode of assembly. Panel (b): The stellar mass surface density of RXCJ2248-ID (orange circle) is significantly elevated relative to measurements from young stellar clusters in the local Universe \citep[light red circles][]{Brown2021}. This regime of increased stellar mass density is shared by individual compact star clusters at high redshift, such as those in the sunrise arc \citep[blue diamonds][]{Vanzella2023}, the sunburst arc \citep[red pentagons][]{Vanzella2022}, or of massive young star clusters in the Cosmic Gems arc \citep[green squares;][]{Adamo2024}.} 
     \label{fig:sizes}
\end{figure*}

\subsection{Emission and absorption line measurements}
\label{sec:measurements}
We implemented multiple fitting procedures to measure absorption and emission line profiles and fluxes that varied depending on the features that were being examined.
Isolated spectral features were initially fit using a single Gaussian using an initial guess for the centroid based on the systemic redshift measured from strong rest-optical lines (e.g., [OIII], H$\alpha$).
The centroid of the line was allowed to vary within a small wavelength range.
Nearby pairs of lines (e.g., H$\gamma$ and [OIII]$\lambda$4363) were fit using a superposition of Gaussian profiles, which we allowed to vary in centroid but had a fixed spacing in wavelength.
The strengths and widths of these two lines were allowed to vary independently. 
Pairs of lines that are unresolved in the spectra (e.g., [OII]) were described by a single Gaussian.
In some special cases, such as in [NIII]$\lambda1750$, additional components were added to the fitting to match the number of multiplets. In each of these cases the spacing of the components was fixed.

In some cases, a single component was not sufficient to describe the observed spectral features (e.g., lines with a broad component; see Sections~\ref{sec:uvlines} and \ref{sec:optlines1} below).
In these instances, we add an additional component to the fitting such that the broad and narrow components were fit simultaneously.
The centroid of both components were allowed to vary independently.
The continuum level was estimated at line center by fitting a linear relation to the spectrum on either side of each line while excluding the emission or absorption line components.
The full width at half-maximum (FWHM) of each noted line was measured from the Gaussian fitting detailed above, and has had the instrumental resolution at the observed wavelength subtracted off in quadrature. 
For lines where the fitting yields widths narrower than the instrument resolution, we provide FWHM upper limits set at the instrument resolution itself.
We obtained uncertainties on each property (total flux, centroid, FWHM, and EW) by first creating 5000 mock spectra that were derived by perturbing the observed spectrum by its corresponding error spectrum.
We repeated the above fitting procedure on each of the mock spectra, and assign uncertainties based on the inner 68th percentile of the recovered distributions of each property.

Using the final combined spectrum we check for consistency in the wavelength solution across the observed spectrum.
We compare the systemic redshift measured using emission lines within each of the NIRSpec grating setups (see Section~\ref{sec:results} for details). 
We measure systemic redshifts of $z_{\rm sys}^{\rm G140M}=6.1060\pm0.0008$, $z_{\rm sys}^{\rm G235M}=6.1056\pm0.0005$, and $z_{\rm sys}^{\rm G395M}=6.1056\pm0.0004$ from the G140M/F100LP, G235M/F170LP, and G395M/F290LP spectra, respectively.
These values are in excellent agreement, suggesting that the spectrum has a very accurate wavelength solution across the gratings.
When used in combination, we find a systemic redshift measured using all of the available lines is $z_{\rm sys}=6.1057\pm0.0006$.

\section{Broadband photometric properties}
\label{sec:stars}
Prior to {\it JWST}, the basic physical properties of RXCJ2248-ID (i.e., stellar mass, light-weighted age) were largely unconstrained as its rest-optical photometry from {\it Spitzer}/IRAC photometry was heavily confused by neighboring galaxies. 
Before discussing the emission lines in the new spectrum of RXCJ2248-ID, we first consider the properties 
implied by broadband measurements of the continuum between 
rest-UV and rest-optical. 
While the recent NIRCam imaging reveals multiple distinct components, we first consider the most luminous clump that dominates the overall emission. We then return to consider contributions from the fainter clumps.

We infer integrated galaxy properties using two methods. 
First, we use our NIRSpec spectrum to synthesize broadband flux densities using the NIRCam filter set that is now commonly utilized by previous analyses (i.e., F090W, F115W, F150W, F200W, F277W, F335M, F356W, F410M, F444W).
The resulting broadband fluxes are displayed in the inset panel of Figure~\ref{fig:mainsequence}.
We calculate the UV slope, $\beta$ (where $f_{\lambda}\propto\lambda^{\beta}$) by fitting a power-law to the synthesized F115W, F150W, and F200W fluxes.
This procedure demonstrates that the UV continuum is very blue, with a power-law slope of $\beta=-2.72$.
As a comparison, we further consider photometry from existing NIRCam imaging in the RXC J2248-4431 field (PID 1840; PI: Alvarez-Marquez, see \S\ref{sec:observations}).
In comparing these measurements, we use photometry from  ID1 as it is the most magnified and brightest of the images.
We find agreement between the UV continuum slope measured from these two methods; the NIRCam photometry yields a similar UV slope of $\beta=-2.60$. The synthesized broadband SED further shows the color excesses that associated with extremely strong rest-optical lines.
As is clear in Figure~\ref{fig:mainsequence}, 
we find a flux excess 
in F410M from $\rm [OIII]+H\beta$ ($m_{\rm F410M}-m_{\rm F356W}=1.80$) 
and in F444W from H$\alpha$ 
($m_{\rm F410M}-m_{\rm F444W}=0.95$).
The $m_{\rm F300M}-m_{\rm F444W}$ excess measured from NIRCam ($1$ AB mag) implies a  consistent H$\alpha$ strength to that described above. However none of existing NIRCam data probe the $\rm [OIII]+H\beta$ line.

We now consider the physical properties implied by the photometry. Here we use the BayEsian Analysis of GaLaxy sEds \citep[\texttt{BEAGLE}][]{Chevallard2016} software package.
These models implement updated templates of \citet{Bruzual2003}, which include nebular emission using the prescription detailed in \citet{Gutkin2016}.
We assume a constant star-formation history (CSFH) with a minimum age of 1 Myr, and adopt a uniform prior on metallicity (ionization parameter) within the range $\log(Z/Z_{\odot})\in [-2.24, 0.3]$ ($\log(U)\in [-4, -1]$).
The redshift of the SED models is fixed to the spectroscopic redshift, and we assume an attenuation law following the SMC extinction curve \citep{Pei1992}. 
The assumption of a CSFH is aimed at understanding the most recently formed stellar population, which dominates the emission in the rest-UV and rest-optical.
This stellar population best representing the source of the spectrum which comprises the focus of this analysis.
It is conceivable that an underlying older stellar population may be present, while contributing minimally to the observed SED.
In this case, some of the inferred properties, such as the stellar mass, may be underestimated considerably \citep[e.g.,][]{Whitler2023a}.
Furthermore, we do not consider contributions from an AGN in our modelling. We discuss the possibility that an AGN impacts the emission from RXCJ2248-ID in \S~\ref{sec:derived} and \S~\ref{sec:disc}, and this will be considered in more detail by Plat et al. 2024, (in prep).

The best-fit SED implies that RXCJ2248-ID is undergoing a period of rapid growth.
Properties inferred from this modeling are presented in Table~\ref{tab:properties}.
We find a stellar mass $\log(\rm M/M_{\odot})=8.05^{+0.17}_{-0.15}$ and star-formation rate (SFR) of $\log(\rm SFR/M_{\odot}yr^{-1})=1.8^{+0.2}_{-0.2}$.
These results imply a high specific-SFR (sSFR) of $560~\rm Gyr^{-1}$, which is more than 1 dex above typical values found at this redshift \citep[e.g.,][]{Speagle2014,Topping2022,Popesso2023}. 
Correspondingly, we find an extremely young age of just $1.8^{+0.7}_{-0.4}$ Myr.
This is further supported by the large emission-line flux excesses; we find $\rm [OIII]+H\beta$ and H$\alpha$ EWs of $3706^{+378}_{-505}\angstrom{}$ and $1088^{+203}_{-189}\angstrom{}$, respectively.
The very blue UV continuum slope requires that the impact of dust be minimal. We infer an effective V-band optical depth of $\tau_{\rm V}=0.02^{+0.03}_{-0.02}$, which is consistent with no attenuation.

We combine properties derived from the best-fit SED with the sizes measured in  Section~\ref{sec:observations}.
Figure~\ref{fig:sizes}a shows RXCJ2248-ID on the size versus UV luminosity relation.
We compare these values to measurements from galaxies in the literature, comprising galaxies and individual components of galaxies in the field and behind lensing clusters \citep{Bouwens2021, Vanzella2022, Chen2023, Atek2023, Tacchella2023, Vanzella2023}.
This comparison indicates that RXCJ2248-ID is much more compact than other systems that share its \muv{}, indicating we are seeing a component of a galaxy with an extremely high density 
of UV photons. A similar offset toward 
high UV densities is seen for GN-z11 \citep{Tacchella2023} and among several  individual star-forming knots and protoglobular clusters \citep[purple squares in Fig.~\ref{fig:sizes}a][]{Vanzella2022, Vanzella2023}. 
The similarity in UV luminosity density between RXCJ2248-ID and these systems may imply they coexist in a similar evolutionary phase. We constrain the star-formation rate density ($\Sigma_{\rm SFR}\equiv\frac{\rm (SFR /M_{\odot}yr^{-1})}{2\pi (r_e/\rm kpc)^2}$) and stellar mass surface density ($\Sigma_{\rm M_*}$) based on the inferred SFR and stellar mass, in addition to the lower limit to the size of the dominant component.  
The results indicate that RXCJ2248-ID has a SFR surface density in excess of $10400~\rm M_{\odot}/yr/kpc^2$ and a stellar mass surface density greater than $3.6\times10^{4}~\rm M_{\odot}/pc^2$ (see Figure~\ref{fig:sizes}b). 
This limit on the SFR surface density exceeds the theoretical Eddington limit for star formation \citep[e.g.,][]{Hopkins2010}, which can be exceeded on very short time scales.
Furthermore, in the case of low metallicities and low dust content, the opacity setting this limit is reduced, which may allow for higher $\Sigma_{\rm SFR}$ to exist.
The other systems noted in Figure~\ref{fig:sizes}a have $\Sigma_{\rm SFR}$ approaching, yet not exceeding this limit; GN-z11 has an SFR surface density in excess of $>820\rm M_{\odot}/yr/kpc^2$ \citet{Tacchella2023}, while the systems in
\citet{Vanzella2022,Vanzella2023} range from $200-4000\rm M_{\odot}/yr/kpc^2$.
This limit suggests that RXCJ2248-ID has among the highest density of star formation 
known at high redshift.

To place these results in broader context, 
we now compare the inferred stellar mass and SFR of RXCJ2248-ID to the galaxy population in the reionization era.
Figure~\ref{fig:mainsequence} compares RXCJ2248-ID to $z\simeq6-9$ star-forming galaxies form \citet{Endsley2023b}, as well as the star forming main sequence derived from \citet{Speagle2014} and calculated at $z=6$.
The consistency between the \citet{Endsley2023} galaxies and the \citet{Speagle2014} main sequence indicates that they comprise a representative sample.
As expected, RXCJ2248-ID is significantly above ($+1.5$ dex) this relation, and has among the highest SFR at fixed stellar mass of galaxies in this epoch.
We additionally compare to GN-z11, which previous analyses have indicated is rapidly assembling \citep[e.g.,][]{Tacchella2023}.\footnote{We adopt photometry from \citet{Tacchella2023} and recompute the best-fit SED using our modelling setup for consistency.}
While we treat the emission from GN-z11 as being of stellar origin for this comparison, recent analyses have suggested that an AGN is present \citep[e.g.,][]{Maiolino2023}.
Similar to RXCJ2248-ID, the best-fit SED of GN-z11 indicates that it sits at a higher SFR for fixed stellar mass compared to galaxies up to $z\simeq9$, albeit by a smaller amount (0.5 dex).
If we interpret these results as both objects existing during a burst phase, we may expect that the spectra of both objects imply similar conditions.

The broadband SED modelling further constrains the ionizing properties of RXCJ2248-ID.
Based on the young age, high sSFR, and very strong emission-line EWs, we expect RXCJ2248-ID to be an efficient producer of ionizing photons.
We compute the ionizing photon production efficiency ($\xi_{\rm ion}$) that has not been corrected for contributions from nebular continuum emission or dust attenuation.
The results from the SED modelling produces a value of $\log(\rm \xi_{\rm ion}/erg^{-1}~Hz)=25.94^{+0.11}_{-0.07}$ for this object.
This is 0.2 dex higher than the median $\xi_{\rm ion}$ found for objects at this UV luminosity and redshift, and is in the highest 16th percentile of such objects \citep[e.g.,][]{Endsley2023b}.
As a result, the spectroscopy of RXCJ2248-ID offers the opportunity to explore the properties of one of the most efficient ionizing agents in this epoch.

The extremely blue UV slope alone can begin to inform on the ISM properties of RXCJ2248-ID.
In addition to the absence of dust attenuation, the UV slope of $\beta=-2.72$ is in the regime where the precise shape of the nebular continuum is critical.
This is particularly important in RXCJ2248-ID, as the inferred properties imply that as much as half of the UV continuum could be of nebular origin \citep[e.g.,][]{Topping2022b}.
The addition of nebular emission to the stellar continuum can significantly redden the spectrum.
One conceivable process that can allow for such blue colors is the escape of ionizing radiation \citep[e.g.,][]{Robertson2010, Zackrisson2013}.
Additionally, high electron densities can lessen the emission from the two-photon process, which dominates the nebular continuum at 1216-2500\angstrom{} in the rest frame \citep{Osterbrock}.
In both scenarios, the nebular continuum is suppressed, and the observed emission increasingly probes the underlying blue stellar continuum.
Either one, or a combination of these processes may be at play in RXCJ2248-ID; we utilize the spectroscopic constraints to explore these possibilities in Section~\ref{sec:derived}.

Thus far, our analysis has focused on the primary component of RXCJ2248-ID. As mentioned in Section \S~\ref{sec:observations}, there is an additional clump separated by $\sim220$ pc that comprises $30\%$ of the total flux density in the F115W filter. Here we consider the SED and implied properties of this fainter star forming component. This secondary component falls entirely within the NIRSpec shutter targeting RXCJ2248-ID1 (see Figure~\ref{fig:images}), and is $0.8$ mag fainter than the primary component in the rest-frame UV.
In addition to the difference in overall luminosity, the rest-UV and optical colors are distinct between the two components.
For the primary component, we derived a UV continuum slope of $\beta=-2.72$ and an $\rm [OIII]+H\beta$ EW of $3706^{+378}_{-505}\angstrom{}$, while this secondary component is characterized by a UV slope and EW of $\beta=-2.20$ and $910^{+530}_{-290}\angstrom{}$, respectively.
Using the best-fit BEAGLE SED, we infer a stellar mass of $\log(\rm M/M_{\odot})=7.5^{+0.1}_{-0.1}$ of the secondary component, which is a factor of three lower than the primary component.
In addition, we find a lower SFR (sSFR) for this component of $\log(\rm SFR/M_{\odot}yr^{-1})=0.4^{+0.3}_{-0.2}$ ($79~\rm Gyr^{-1}$), and a corresponding galaxy age of $t_{\rm age}^{\rm CSFH}=13^{+12}_{-8}$ Myr.
While the properties place this secondary component at a higher SFR at fixed stellar mass relative to most objects during this epoch, it is a far less extreme example when compared to the primary component of RXCJ2248-ID.
As such, it is not surprising that the emission lines are dominated by the primary component.

\begin{figure*}
    \centering
     \includegraphics[width=1.0\linewidth]{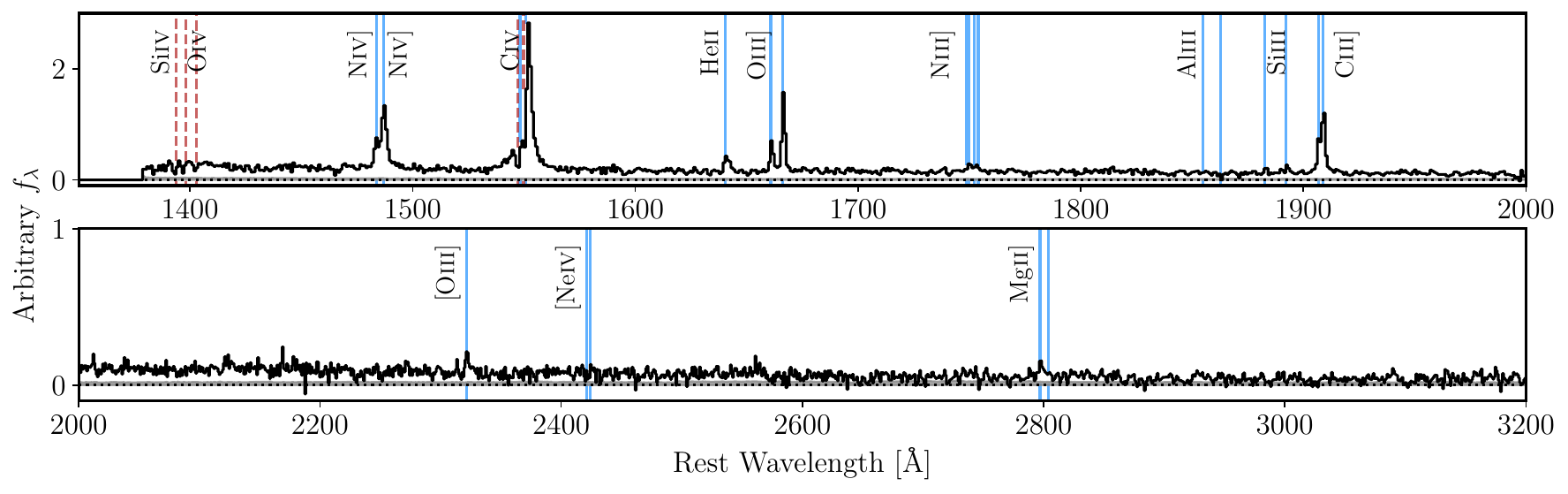}
     \caption{Final 1D spectrum of RXCJ2248 lensed object at $z_{\rm spec}=6.11$, providing full wavelength coverage from rest-frame 1370\angstrom{} to  3200\angstrom{}. This spectrum is the result of stacking the three multiply-lensed images for which spectra were obtained (see Section~\ref{sec:data}).  We display the error spectrum as a grey shaded region. Each emission line that is detected at $>3\sigma$ is identified by a vertical blue line. Absorption features that are significantly detected are indicated by a vertical dashed red line.} 
     \label{fig:uvspectrum}
\end{figure*}

%
%
%
%
\section{JWST Spectroscopic Properties}
\label{sec:results}

The \jwst{}/NIRSpec data of RXCJ2248-ID provide more 
insight into the ionizing sources and gas properties 
linked to strong CIV emission in the reionization era. 
Here we analyze the stack of the spectra from the three lensed images (see \S\ref{sec:two}), yielding a high S/N spectrum with continuous wavelength coverage from 9790\angstrom--52000\angstrom{}. Using the systemic redshift of RXCJ2248-ID ($z=6.1057$), this 
 corresponds to a rest-frame coverage spanning 1378--7320\angstrom{}, allowing constraints on many rest-UV and rest-optical lines.
In this section, we present observed FWHM, that have not been corrected for instrumental effects unless otherwise noted. For these measurements from the $R\sim1000$ spectra, we consider lines with widths of $<300\rm km/s$ to be unresolved.
\S~\ref{sec:uvlines} presents the rest-UV spectrum and measurements, while the rest-optical spectrum is presented in \S~\ref{sec:optlines1}.
We use these results to constraints gas-phase properties in \S~\ref{sec:derived}.

\subsection{Rest-frame ultraviolet emission and absorption lines}
\label{sec:uvlines}

\begin{table}
\caption{Rest-UV line centroids, line fluxes, equivalent widths, and FWHM. The line FWHM are the measured value, and have not been corrected for instrumental effects. Upper limits are provided at the $3\sigma$ level. 
\newline $^{\rm a}$ The CIV measurement is only provided as a total flux and equivalent width.
\newline $^{\rm b}$ Indicates blended lines where the FWHM is not well defined.}
\begin{center}
\renewcommand{\arraystretch}{1.2}
\begin{tabular}{ccccc}
\toprule
Line & $\lambda_{\rm obs}$  & Flux  & EW  &FWHM \\
 &$\rm \angstrom{}$ & $\rm 10^{-19}erg/s/cm^2$ &\angstrom{} & km/s\\
\midrule

$\rm NIV]\rm \lambda1483$& 10542.0 & $8.6^{+0.4}_{-0.4}$ & $5.1^{+0.3}_{-0.2}$ & $360^{+103 }_{ -90 }$\\
$\rm NIV]\rm \lambda1486$&10567.9 & $20.8^{+0.6}_{-0.6}$ & $12.2^{+0.4}_{-0.3}$ & $360^{+103 }_{ -90 }$\\
$\rm CIV\rm \lambda1549$&11026.3 & $61.7^{+0.9}_{-0.8}$$^{a}$ & $34.1^{+0.8}_{-0.7}$$^{a}$ & $-^{\rm b}$\\
$\rm HeII\rm \lambda1640$&11657.3 & $5.9^{+0.4}_{-0.4}$ & $5.6^{+0.4}_{-0.4}$ & $428^{+102 }_{ -118 }$\\
$\rm OIII]\rm \lambda1660$&11802.0 &  $6.3^{+0.2}_{-0.2}$ & $5.1^{+0.2}_{-0.2}$ & $292^{+75 }_{ -46 }$\\
$\rm OIII]\rm \lambda1666$&11839.9 &  $16.8^{+0.3}_{-0.3}$ & $13.7^{+0.3}_{-0.2}$ & $292^{+75 }_{ -46 }$\\
$\rm NIII\rm \lambda1750$&12434.4 &  $4.5^{+0.4}_{-0.4}$& $4.3^{+0.5}_{-0.4}$ & $-^{\rm b}$\\
$\rm AlIII\rm \lambda1854$&  $-$ & $<1.4$ &  $<2.0$&  $-$\\
$\rm AlIII\rm \lambda1863$&  $-$ & $<1.4$ &  $<2.0$&  $-$\\
$\rm SiIII]\rm \lambda1883$&13380.8 &  $1.7^{+0.4}_{-0.4}$& $2.0^{+0.4}_{-0.4}$ & $278^{+124}_{-94}$\\
$\rm SiIII]\rm \lambda1892$&13446.4 &  $3.0^{+0.4}_{-0.4}$& $3.9^{+0.7}_{-0.6}$ & $278^{+124}_{-94}$\\
$\rm [CIII]\rm \lambda1907$&13547.9 &  $6.9^{+0.3}_{-0.3}$ & $6.7^{+0.3}_{-0.3}$ & $266^{+73 }_{ -55 }$\\
$\rm CIII]\rm \lambda1909$&13563.3 & $15.3^{+0.4}_{-0.3}$ & $15.0^{+0.3}_{-0.3}$ & $266^{+73 }_{ -55 }$\\
$\rm [OIII]\rm \lambda2321$&16496.3 &  $2.8^{+0.5}_{-0.4}$ & $4.8^{+0.8}_{-0.6}$ & $346^{+84 }_{ -77 }$\\
$\rm MgII\rm \lambda2800$ &19875.4 & $2.8^{+1.1}_{-0.8}$ & $7.4^{+2.8}_{-2.0}$ & $312^{+145}_{-126}$\\
\bottomrule
\end{tabular}
\end{center}
\label{tab:UVlines}
\end{table}

The NIRSpec data presented in this section build on earlier spectroscopy of RXCJ2248-ID, which we briefly summarize here for context. A deep Magellan/FIRE spectrum revealed a significant 
emission feature (S/N$=6.3$) attributed to CIV$\lambda1550$ at an observed-frame wavelength of 11023.8\angstrom{} \citep{Mainali2017}. No corresponding CIV$\lambda1548$ emission was detected in the FIRE spectrum, suggesting contamination 
by a nearby OH sky line. \citet{Schmidt2017} reported a comparable
(3-5$\sigma$) CIV detection in HST WFC3/IR grism observations. 
\citet{Mainali2017} also report detections in OIII]$\lambda1660$ and OIII]$\lambda1666$ at a significance of $2.9\sigma$ and $4.5\sigma$, respectively. The remaining UV lines (i.e., He II, CIII]) were not detected with FIRE or the WFC3/IR grism.
The new NIRSpec data provide a much deeper view of the rest-UV of RXC J2248-ID (Figure \ref{fig:uvspectrum}), revealing numerous emission lines and several absorption lines. 
We describe these spectral features below. 

\subsubsection{CIV and He II}

\begin{figure}
    \centering
     \includegraphics[width=1.0\linewidth]{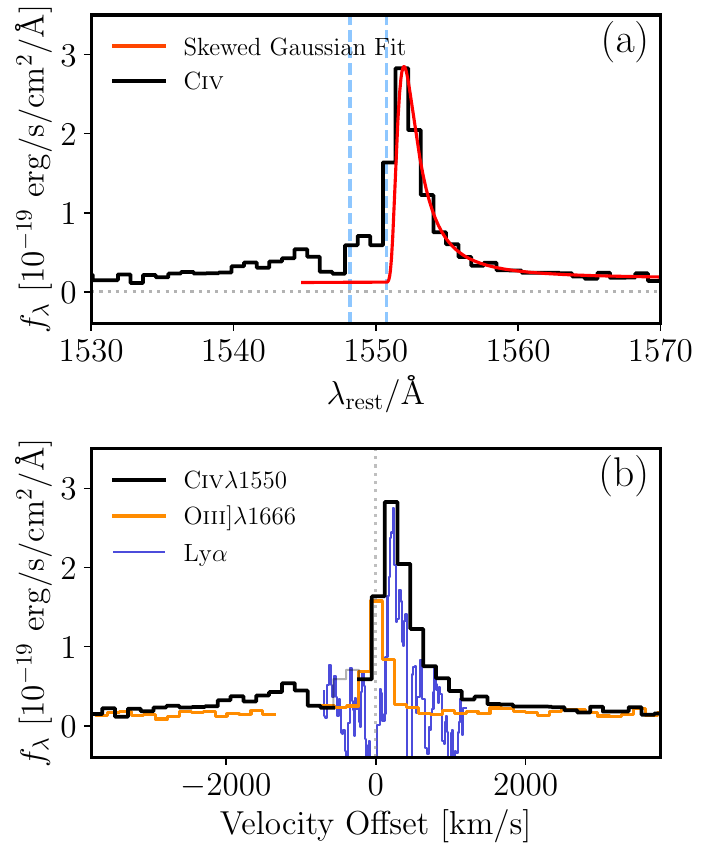}
     \caption{CIV$\lambda\lambda1548,1550$ profile in RXCJ2248-ID. Panel (a) displays the CIV profile overlayed by a skewed Gaussian profile fit. The vertical blue lines indicate the rest-frame wavelengths of the CIV doublet members. In panel (b) we show the velocity structures of CIV$\lambda\lambda1548,1550$. We display the CIV profile as a black histogram, and overlay the OIII]$\lambda1666$ profile as an orange histogram. For clarity, we mask out the OIII]$\lambda1660$ component, and display the CIV$\lambda1548$ component at a reduced opacity.  In addition, we display the Ly$\alpha$ profile presented by \citet{Mainali2017} as the blue histogram. We normalize the peak of the Ly$\alpha$ profile to that of the CIV emission line.}
     \label{fig:broaduv}
\end{figure}

\begin{figure*}
    \centering
     \includegraphics[width=1.0\linewidth]{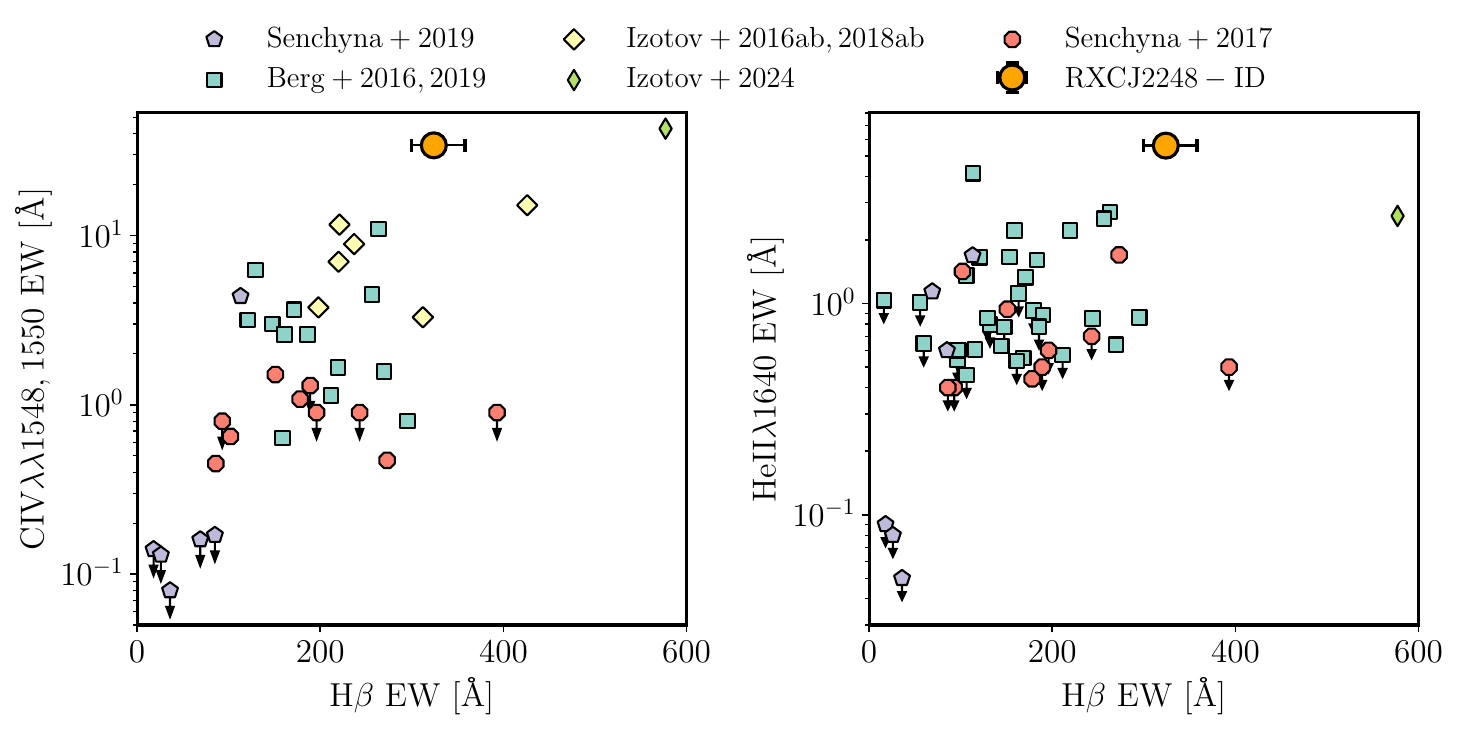}
     \caption{Rest-frame equivalent widths of CIV (panel a) and HeII (panel b) as a function of H$\beta$ EW (as discussed in Section~\ref{sec:optlines1}). We compare the measurements of RXCJ2248-ID (orange point) to measurements of low-metallicity galaxies at low redshift and in the local Universe, including \citet{Senchyna2019} (violet pentagons), \citet{Izotov2016a, Izotov2016b, Izotov2018, Izotov2018b} (yellow diamonds), \citet{Berg2016, Berg2019} (turquoise squares),  \citet{Senchyna2017} (red octagons), and \citet{Izotov2024} (green diamond). RXCJ2248-ID continues the trend seen in the comparison samples, in that the highest CIV and HeII EWs exist among the largest H$\beta$ EWs. However, the HeII EW in RXCJ2248-ID exceed all of the lower-redshift objects, and it has the second-highest CIV EW among these comparisons. This suggests that the observed properties of RXCJ2248-ID may be unique to the reionization era.} 
     \label{fig:uvews}
\end{figure*}

Our new {\it JWST} spectrum confirms the  bright CIV emission detected in earlier 
spectroscopy (Figure~\ref{fig:broaduv}). The line reaches its peak 
flux at 11026.3~\angstrom{}  (rest-frame 1551.75\angstrom{}) and has a S/N of $77$. We note that this is more than 10-20$\times$ greater than the S/N of the Magellan and {\it HST} detections that motivated this study. From integrating over all of the CIV emission that lies above the continuum, we derive a total rest-frame EW of $34.1^{+0.8}_{-0.7}$\angstrom{}.  Such strong 
line emission is rarely seen in  very low-metallicity dwarf galaxies in the local Universe (see compilation in Figure~\ref{fig:uvews}), with only one recently-detected system exceeding the measured value in RXCJ2248-ID \citep{Izotov2024}.
The line appears dominated by a relatively narrow  ($\rm FWHM=231^{+64}_{-29}~\rm km/s$) component, with the rest-frame wavelength suggesting that we are primarily detecting emission from CIV$\lambda1550$. The peak flux  of the line is redshifted by 200 km/s, and the line additionally shows a clear asymmetry with an extended 
tail of red-side emission. Both of these 
characteristics are consistent with expectations for resonant transfer of CIV photons through outflowing gas \citep{Berg2019,Berg2019b,Senchyna2022}. We detect only a weak component associated with the blue side of the doublet (also redshifted by 200 km/s from systemic). The absence of strong CIV$\lambda$1548 emission is consistent with 
earlier ground-based measurements, and likely 
points to resonant scattering significantly 
attenuating the blue component of the doublet. 
We will discuss this in more detail below.

A closer look at the spectrum (Figure~\ref{fig:broaduv}) reveals additional complexities in the CIV profile. In addition to the emission described above, we observe blueshifted ($202\pm33$ km/s) absorption lines (see Section~\ref{sec:abslines} for a more comprehensive discussion). The absorption 
is strongest for the CIV$\lambda$1548 component, but we also detect a hint of absorption just blueward of CIV$\lambda$1550. 
The absorption lines confirm the presence of the CIV outflowing gas which is modulating the emission spectrum. 
The strongest peak in the CIV profile appears asymmetric, with emission clearly extending to the red of the CIV$\lambda 1550$ component, and is reminiscent of resonant Ly$\alpha$ profiles.
To characterize the profile shape, we fit a skewed Gaussian to the CIV emission redward of the CIV$\lambda1550$ peak (see Figure~\ref{fig:broaduv}a).
This fitted profile peaks at a velocity of $+240$ km/s, and extends in emission to velocities of $+1100$ km/s from systemic. 
The CIV emission profile also displays a weak blue bump that peaks at a velocity of $-590\pm170$ km/s relative to the vacuum wavelength of CIV$\lambda1548$.
This blue bump has a greater velocity offset than the red peak, as is commonly seen in double-peaked Ly$\alpha$ emitters \citep[e.g.,][]{Hashimoto2015,Verhamme2015,Verhamme2018,Furtak2022}.
Such features of resonant line profiles arise naturally through the processing of emitted photons through the outflowing gas. We note that the [OIII] and H$\alpha$ profiles (discussed in Section~\ref{sec:o3ha} below) show broad emission components (in addition to dominant narrow components) with FWHM in the range $600-1200$ km/s. 
These broad components provide a consistent physical picture to the CIV profile, whereby both are best explained if  there is warm outflowing gas with speeds that reach up to 1000 km/s (with a dominant component at $\simeq 200$ km/s).

The HeII $\lambda1640$ recombination line provides another probe of the radiation field in RXCJ2248-ID.  Whereas the earlier observations were not sensitive enough to 
detect HeII, the \jwst{} spectra are able to recover the line 
at a wavelength of 11657.3\angstrom{} ($\rm S/N=14$). We measure a total flux of $5.9\pm0.4\times10^{-19}~\rm erg/s/cm^2$, which is consistent with the upper limit of $<15\times10^{-19}~\rm erg/s/cm^2$ reported by \citet{Mainali2017}.
We present the He II $\lambda1640$ EW of RXCJ2248-ID ($5.6\pm0.4$\angstrom{}) in comparison to star forming galaxies at lower redshift and the local Universe in Figure~\ref{fig:uvews}.
RXCJ2248-ID has the largest He II EW of any of these comparison objects, including those with similar H$\beta$ EW. Clearly RXCJ2248-ID is distinct 
from the vast majority of nearby metal poor 
galaxies, such that RXCJ2248-ID has a radiation field with a 
large output of 54 eV photons. The measured He II FWHM is $428^{+102}_{-118}~\rm km/s$. While this is slightly larger than we find for the non-resonant OIII] emission line ($292$ km/s), it is still consistent 
with the limits of the unresolved lines after correcting for instrumental resolution ($223^{+37}_{-35}~\rm km/s$). As a result, we find no clear 
evidence for broadening that may be associated 
with stellar winds or AGN \citep[e.g.,][]{Senchyna2017, Saxena2020}. 
We will come back to discuss implications of the C IV/He II flux ratio in Section~\ref{sec:uvlineratios}.

\subsubsection{Nitrogen and aluminum emission lines}

Recent analyses have identified a small number of objects at $z>6$ that display emission in NIV]$\lambda\lambda1483,1486$ \citep[e.g.,][]{Larson2023, Bunker2023, Isobe2023, Marques-Chaves2023}.
The presence of this line has led to claims of enhanced nitrogen abundances ($\log(\rm N/O)\gtrsim -0.5$ to $0.5$) in these early systems \citep[e.g.,][]{Senchyna2023, Isobe2023, Cameron2023}, corresponding to super-solar nitrogen content \citep[where $\log(\rm N/O)_{\odot}=-0.86$;][]{Asplund2009}.
Examples of this line are exceedingly rare among star-forming galaxies at lower redshifts, and tend to only exist in a small subset of nitrogen-enhanced dwarf galaxies selected to mimic the stellar and gas-phase properties of high-redshift systems \citep[e.g.,][]{Berg2022, Senchyna2022}.
In contrast, NIV]$\lambda\lambda1483,1486$ is commonly present among AGN \citep[e.g.,][]{Hainline2011, Feltre2016}.

The {\it JWST} spectrum of RXCJ2248-ID shows a clear detection of NIV]$\lambda1486$ (S/N$=42$) and NIV]$\lambda1483$ (S/N$=17$).
The two lines lie at the systemic redshift of the galaxy, and peak at an observed-frame wavelength of 10542.3\angstrom{} (10567.0\angstrom{}) for $\rm NIV]\lambda1483$ ($\rm NIV]\lambda1486$).
The detection of both lines enables us to constrain a line ratio of $f_{\rm NIV]\lambda1483}/f_{\rm NIV]\lambda1486}=0.41^{+0.05}_{-0.05}$, and total EW of $29.4^{+0.7}_{-0.7}~$\angstrom{}.
We note that in contrast, the NIV] emission from GN-z11 is almost entirely contained within the NIV]$\lambda1486$ line (S/N=7 in the R=1000 spectrum), enabling only an upper limit on the doublet ratio (1-sigma) of $f_{\rm NIV]\lambda1483}/f_{\rm NIV]\lambda1486}<0.15$ \citep{Bunker2023, Maiolino2023, Senchyna2023}. 
The density of the NIV]-emitting gas plays a key role in setting this doublet ratio. We return to explore the gas-phase properties in detail in Section~\ref{sec:abundances}.

We now consider NIII]$\lambda1750$, which is detected as a weak line in RXCJ2248-ID at a S/N of $5.6$.
This feature comprises an emission quintuplet\footnote{The NIII] quintuplet is composed of emission lines with rest-frame wavelengths of 1746.82~\angstrom{}, 1748.64~\angstrom{}, 1749.67~\angstrom{}, 1752.16~\angstrom{}, and 1753.99~\angstrom{}.}
which is challenging to disentangle in our data due to the spectral resolution and relatively low flux in the line.
The observed peak of the line is present at $12434.4~$\angstrom{}, which at the systemic redshift corresponds to the NIII]$\lambda1750$ line which is typically is the brightest of the complex.
The presence of both NIV] and NIII] emission in the rest-UV is a clear indication for a significant  abundance of nitrogen in RXCJ2248-ID, which is explored in Section~\ref{sec:abundances}.
We find that NIV] is significantly stronger than NIII], with a flux ratio of $f_{\rm NIV]}/f_{\rm NIII]}=6.5$.
Stronger emission from the higher-ionization line is indicative of very intense ionization conditions, which can be achieved by shocks or AGN \citep[e.g.,][]{Feltre2016, Alarie2019}, but can also result from stellar photoionization in the case of very high ionization parameter (see Plat et al. (in prep) for a detailed discussion).
An opposing constraint is provided by the HeII$\lambda1640$ emission, which is much weaker than NIV] ($f_{\rm NIV]}/f_{\rm HeII})=5.0$). 
Photoionization models suggest this difference in flux is consistent with a sharp drop in the radiation field toward higher energies between 47 and 54 eV \citep[e.g.,][]{Senchyna2023}.
A detailed analysis exploring the implications of the emission-line ratios will be presented by Plat et al. (in prep).

It has been pointed out in the literature that this level of nitrogen-enhancement is similar to that seen in the second generation globular cluster stars \citep{Carretta2009, Pancino2017,Masseron2019}. 
We will come back to this in the discussion section. If the nitrogen-enhancement stems from high-temperature nuclear burning \citep[e.g.,][]{Prantzos2007}, we may expect to see enhancements in aluminum alongside those of nitrogen. Our spectra samples several aluminum lines (AlII$\lambda1670$, AlIII$\lambda\lambda1854,1864$), both of which are commonly seen in absorption in high redshift spectra.
We expect any AlII$\lambda1670$ feature to exist at observed-frame 11866\angstrom{} near the OIII]$\lambda1666$ line, but resolved in the $R\sim1000$ spectrum. 
However no significant line is detected, and we place a $3\sigma$ upper limit on the flux and EW of $<1.6\times10^{-19}~\rm erg/s/cm^2$ and $1.3$\angstrom{}, respectively.
Similarly, neither component of $\rm AlIII\lambda\lambda1854,1862$ show significant detections, and each doublet member is constrained to $<1.4\times10^{-19}~\rm erg/s/cm^2$ (EW$<$$2.0$\angstrom{}) at the $3\sigma$ level.
Future spectroscopy capable of detecting these lines may be required in order to place the aluminum abundance in context of either globular cluster formation or supermassive star activity.

\subsubsection{CIII] and OIII] emission lines}

The [CIII], CIII] $\lambda\lambda1907,1909$ doublet 
is typically one of the strongest UV lines in metal poor dwarfs, but ground-based observations of RXCJ2248-ID 
 only placed an upper limit on the line flux. 
In our NIRSpec observations, we are able to separately identify both components of the doublet. When simultaneously fitting both lines, we find centroids at 13547.9\angstrom{} and 13563.3\angstrom{}, yielding a consistent systemic redshift  ($z_{\rm sys, UV}=6.1057\pm0.0004$) to that measured from other non-resonant UV and optical lines. By separating the two doublet members, we derive a flux ratio of $f_{\rm CIII]\lambda 1909}/f_{\rm [CIII]\lambda 1907}=2.2^{+0.3}_{-0.3}$, and we measure a total EW of $21.7^{+0.7}_{-0.7}$\angstrom{}.
This regime (i.e., $f_{\rm CIII]\lambda 1909}>f_{\rm [CIII]\lambda 1907}$) implies electron densities in excess of $\sim10^4~\rm cm^{-3}$, which we provide more detail on in Section~\ref{sec:density}.
We note that our measurement of the total CIII] flux of $2.2\times10^{-18}~\rm erg/s/cm^2$ is consistent with previous upper limits set by the HST WFC3/IR grism ($<3.6\times10^{-18}~\rm erg/s/cm^2$, \citealt{Mainali2017}).
The observed CIII] emission can be compared with the CIV and HeII lines to provide insight into the ionizing radiation field.
We measure a CIV/CIII] ratio of $2.8^{+0.4}_{-0.3}$ which by itself is consistent with a wide range of both AGN and stellar photoionization models \citep[e.g.,][]{Feltre2016, Gutkin2016}.
However in the latter case, stars with low ($Z<0.1Z_{\odot}$) metallicities are required.
Additional discriminating power is possible when combined with the CIII]/HeII flux ratio. We derive a value of $f_{\rm CIII]}/f_{\rm HeII}=3.8^{+0.2}_{-0.3}$, which in combination with the CIV/CIII] ratio is in much better agreement with models of ionization by stars (see \S~\ref{sec:uvlineratios} below).

The spectrum of RXCJ2248-ID yields detections in both auroral OIII]$\lambda\lambda1660,1666$ doublet members at observed-frame wavelengths of 11802.0\angstrom{} (S/N$=$$17$) and 11839.9\angstrom{} (S/N$=$$38$), respectively.
Due to their strength and relative isolation, these lines provide an excellent confirmation of the systematic redshift in the rest-frame UV.
Using both lines, we derive a systemic redshift of $z_{\rm sys,UV}=6.1061$.
The [OIII]$\lambda2300$ line is detected ($\rm S/N=7$) at $\lambda=16497.6$\angstrom{}.
This line can be used in combination with the OIII]$\lambda\lambda1660,1666$ emission to extract electron temperatures of the gas.
While significantly weaker than the [OIII] emission lines in the rest-optical, using the [OIII]$\lambda2300$ line may be advantageous due to the much closer wavelength spacing to OIII], reducing the impact of uncertainties driven by dust attenuation.
The OIII]$\lambda\lambda1660,1666$ line is also a key component in deriving abundance patterns in the high-ionization phase of the ISM. 
Two key abundances enabled by the OIII] line are the N/O and C/O ratios, which we explore quantitatively in Section~\ref{sec:abundances}.
Finally, we find that the OIII] emission is significantly stronger than HeII$\lambda1640$, with a ratio of $f_{\rm OIII]}/f_{\rm HeII}=3.9^{+0.7}_{-0.8}$. 
This is consistent with the lower limit on this ratio derived in \citet[$f_{\rm OIII]}/f_{\rm HeII}>$$2.9$; ][]{Mainali2017}, which was found to be far above values that requiring an AGN interpretation.

\subsubsection{Si III emission lines}
Enrichment of silicon in the ISM is expected to follow that of oxygen due to their parallel formation pathways. The relative abundance ratio of silicon to oxygen is often constrained in metal poor galaxies via the the detection of the Si III]$\lambda\lambda1883,1892$
doublet. We detect both components of Si III] in RXCJ2248-ID, with the Si III]$\lambda$1883 feature at 13380.8\angstrom{} ($\rm S/N=4.3$; $\rm EW=2.0^{+0.4}_{-0.4}\angstrom{}$) and Si III]$\lambda$1892 feature at 13446.4\angstrom{} ($\rm S/N=6.0$; $\rm EW=3.9^{+0.7}_{-0.6}\angstrom{}$).
The relative flux of the Si III] doublet members is sensitive to the electron density over a range of $10^3$ to $10^6$ $\rm cm^{-3}$, which we will come back to discuss in detail in Section~\ref{sec:density}.
One key process capable of regulating the strength of silicon line emission is the preferential depletion of silicon on to dust grains relative to other elements at a fixed metallicity \citep[e.g.,][]{Garnett1995, Peimbert2010, Steidel2016}.
As a result, the strength of the SiIII] lines can be a sensitive probe of the dust-to-metal mass ratio ($\xi_d$) in the ISM \citep{Gutkin2016}.
At the metallicity we derive for RXCJ2248-ID ($1/20 Z_{\odot}$, see Section~\ref{sec:abundances} below), the measured SiIII]/CIII] ratio ($\log(\rm SiIII] / CIII])=-1.1^{+0.1}_{-0.1}$) requires a value of $\xi_d$ below $0.3$. 
Further more, if we simultaneously consider the CIII]/OIII] ratio, the models of \cite{Gutkin2016} that best reproduce the observed emission lines have a lower dust-to-metal ratio of $\xi_d=0.1$.
This is in broad agreement with the absence of attenuation that affects the rest-optical emission lines (see Section~\ref{sec:optlines1} below) which implies minimal dust content.

\begin{figure}
    \centering
     \includegraphics[width=1.0\linewidth]{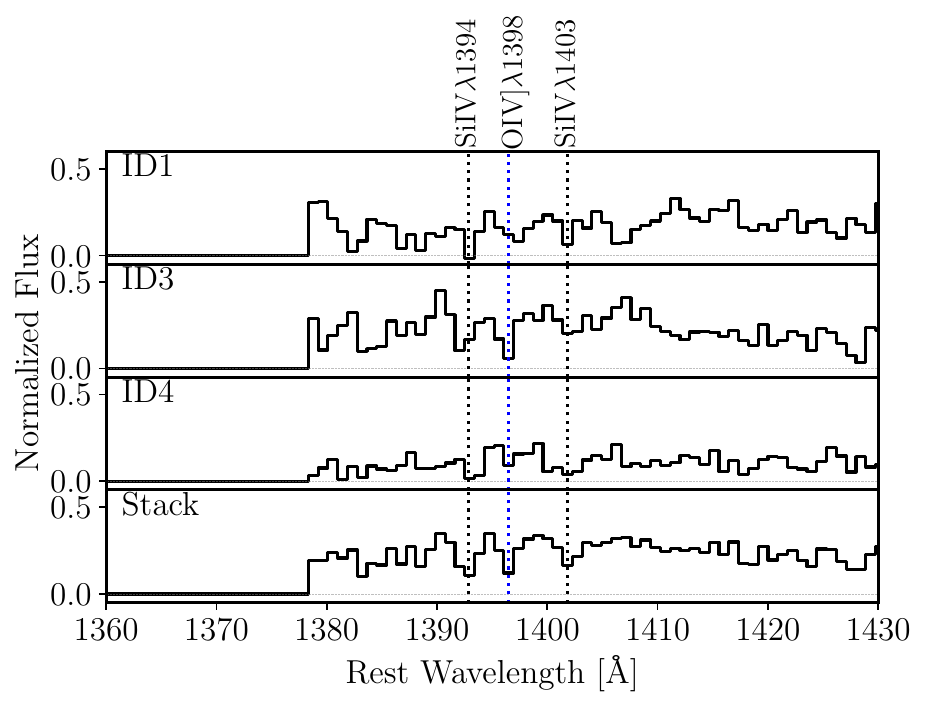}
     \caption{High ionization absorption lines in the spectrum of RXCJ2248-ID. We label the positions of SiIV$\lambda\lambda1394,1403$ and OIV]$\lambda1398$. This figure presents the spectra from the three different lensed images individually (top three panels), in which the absorption lines are visible. In the bottom panel we show the inverse variance-weighted stack of these spectra where the absorption lines are clearly visible. The vertical lines mark the measured locations of the absorption lines which are offset from systemic described in Section~\ref{sec:uvlines}.} 
     \label{fig:HIS}
\end{figure}

\subsubsection{Mg II emission lines}
The MgII$\lambda\lambda2797,2803$ doublet is a sensitive probe of intervening neutral gas in the ISM and CGM \citep[e.g.,][]{Henry2018, Chisholm2020}.
The systemic redshift of RXCJ2248-ID places these lines in the G235M/F170LP coverage at 19869--19912\angstrom{}. 
Within this wavelength range, we detect a single line in emission (S/N$=3.7$) at 19875.4\angstrom{} corresponding to MgII$\lambda2797$, and measure a flux of $2.8^{+1.1}_{-0.8}\times10^{-19}~\rm erg/s/cm^2$ ($\rm EW=7.4^{+2.8}_{-2.0}\angstrom{}$). This line is consistent with being narrow, with a FWHM of $312^{+145}_{-126}~\rm km/s$.
The MgII$\lambda2803$ line is not detected, and we place a $3\sigma$ upper limit on the flux of $<2.4\times10^{-19}~\rm erg/s/cm^2$, which yields a lower limit on the flux ratio of $f_{2797}/f_{2803}>1.2$ at the $3\sigma$ level.
This flux ratio constraint is broadly consistent with the regime where MgII is optically thin, which in turn, implies a low column density (or similarly, low covering fraction) of neutral gas \citep[e.g.,][]{Chisholm2020}. This is consistent with the 
strong Ly$\alpha$ emission in this galaxy \citep{Mainali2017}, and may suggest this source presents conditions for Lyman Continuum leakage. In the case of optically thick Mg II, we expect the line ratio to be unity, such that we likely would have detected both lines in our spectrum. 
Future work offers the potential to build on these 
observations with deep enough spectra to recover 
both components of Mg II, placing more reliable constraints on the neutral gas opacity.

\subsubsection{Insight from absorption lines}
\label{sec:abslines}

We detect the rest-UV continuum at high S/N in our 
stacked spectrum (S/N=$6$ per resolution element), enabling absorption line strengths to be constrained.
Previously, we described narrow C IV absorption (see Figure~\ref{fig:broaduv}), however we also observe absorption in the high-ionization Si IV and O IV lines.
The SiIV$\lambda\lambda1394,1403$ lines are seen at $9899.0\angstrom{}$ and 9962.3\angstrom{}, respectively.
We find another absorption feature at a wavelength of 9925.1\angstrom{}, which is consistent with the expected location of OIV]$\lambda1397$.
While these absorption line features are weak, we find that they are present in each of the three observed lensed images of RXCJ2248-ID, bolstering 
confidence that these are real features.
In Figure~\ref{fig:HIS}, we show the individual spectrum of each image, as well as the final stacked spectrum.
Each of these high-ionization absorption features is blueshifted relative to the systemic redshift.
We measure a velocity offset of $170~\rm km/s$ for Si IV$\lambda\lambda1394,1403$, $142~\rm km/s$ for O IV]$\lambda1397$ (Figure~\ref{fig:HIS}), and $176~\rm km/s$ for C IV$\lambda1548,1550$ (see Figure~\ref{fig:broaduv}).
These velocities are consistent with being driven by star formation \citep[e.g.,][]{Pettini2001, Du2016}, while the energetics of AGN can result in outflows of $>1000~\rm km/s$ \citep[e.g.,][]{Elvis2000, Matthee2023, Maiolino2023}.

The flux ratios of absorption lines (i.e., Si IV$\lambda\lambda1393,1402$) can be used to 
determine if the transition is saturated. 
The EW of Si IV$\lambda1394$ ($-1.1\pm0.3$\angstrom{}) is consistent with being a factor of two greater than that of Si IV$\lambda1402$ ($-0.7\pm0.2$\angstrom{}), matching their relative strengths set by atomic physics \citep{Verner1994}. 
However the doublet members are also broadly consistent (at 1$\sigma$) with having the same EW. In the former case, the transitions would be in the linear regime of the curve-of-growth implying the ionized gas is optically thin.
If instead the doublet members are saturated, the covering fraction of optically thick highly ionized gas implied from the residual flux at line center is $\simeq 40$\%.  A deeper spectrum would more readily distinguish between the optically thin and saturated cases for the absorption lines.

\subsubsection{Insight from UV line ratios}
\label{sec:uvlineratios}
\begin{figure*}
    \centering
     \includegraphics[width=1.0\linewidth]{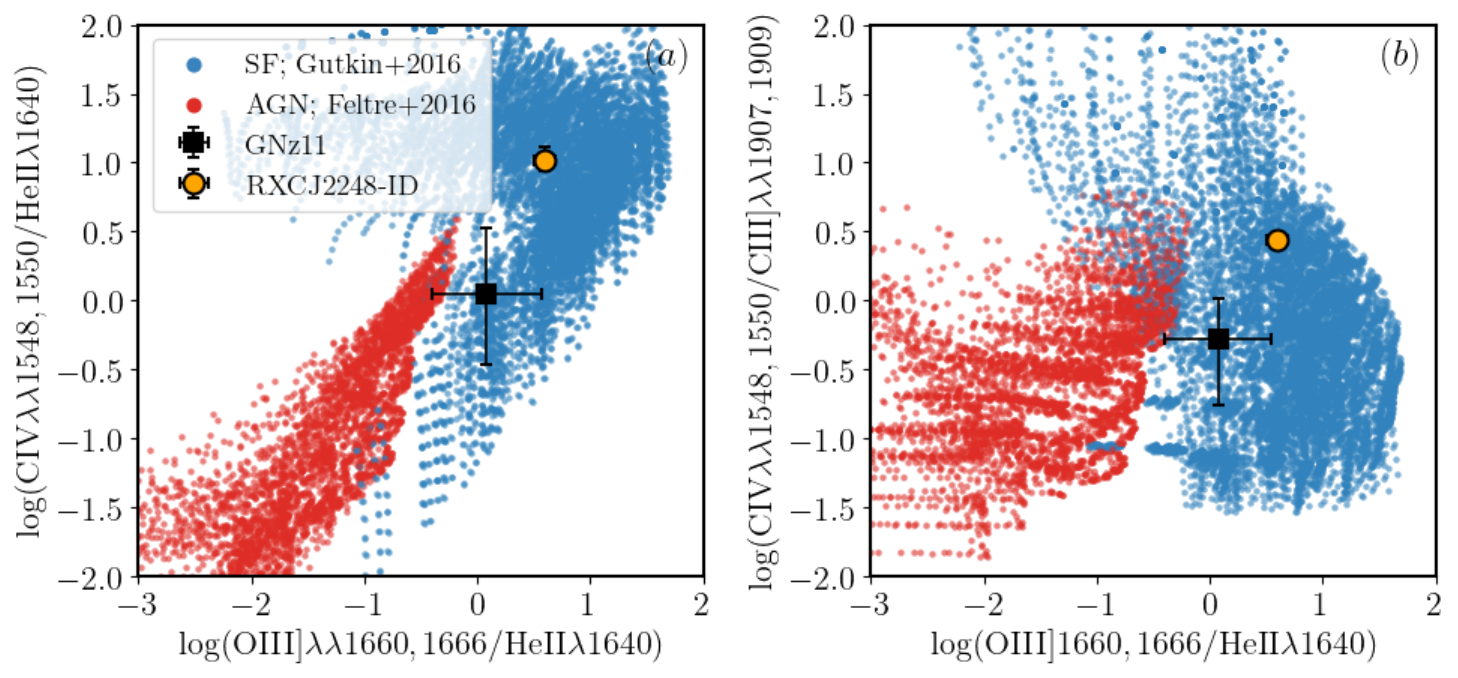}
     \caption{Rest-UV line ratios in the context of photoionization models of star-forming galaxies (blue points; \citealt{Gutkin2016}), and AGN-driven ionization (red points; \citealt{Feltre2016}). We present the CIV/HeII versus OIII]/HeII line ratios in panel (a), and the CIV/CIII] versus OIII]/HeII ratios in panel (b). Line ratios derived for RXCJ2248-ID are displayed as the orange point. For comparison, we place measurements of GN-z11 \citep{Bunker2023} on this diagram indicated by a black marker. This comparison demonstrates that the ionizing spectra of RXCJ2248-ID is consistent with being driven by massive stars, and both have OIII]/HeII and CIV/HeII in excess of that expected from AGN photoionization models.} 
     \label{fig:ionization}
\end{figure*}

We analyze ratios of lines with different ionization energies to characterize the shape of the ionizing spectrum.
Combinations of $\log(\rm C\textsc{iv}/He\textsc{ii}) = 1.02^{+0.08}_{-0.10}$,  $\log(\rm O\textsc{iii}]/He\textsc{ii}) = 0.59^{+0.09}_{-0.10}$, and $\log(\rm C\textsc{iv}/C\textsc{iii}]) = 0.44^{+0.05}_{-0.05}$ offer the potential to discriminate between SF and AGN scenarios.
Figure~\ref{fig:ionization} presents these line ratio alongside models of nebular emission  driven by both narrow-line AGN \citep{Feltre2016} and  star formation \citep{Gutkin2016}.
The star-formation model grid comprises single-star models with solar abundance patterns constructed using stellar evolutionary tracks from \citet{Bressan2012} and \citet{Chen2015}, combined with libraries  of ionizing spectra for OB stars from \citet{Lanz2003,Lanz2007}, and for WR stars from \citet{Hamann2004}.
RXCJ2248-ID sits comfortably within the star-forming models, having $\log(\rm C\textsc{iv}/He\textsc{ii})$ and $\log(\rm O\textsc{iii}]/He\textsc{ii})$ that are greater than the predictions from narrow-line AGN.
The relatively high line ratios may signify a substantial drop-off of photon production at energies above the HeII ionizing edge (54 eV).
While the presence of strong nebular HeII indicates that the ionizing spectrum does extend above 54 eV, we may expect a type II AGN to produce stronger He II.

The UV line ratios of RXCJ2248-ID are somewhat distinct from those 
of GNz11, another source with strong high ionization emission lines \citep{Bunker2023}.
In particular, we note that the GN-z11 spectrum has $\log(\rm C\textsc{iv}/He\textsc{ii})$ and $\log(\rm O\textsc{iii}]/He\textsc{ii})$ values that are below that of RXCJ2248-ID. This may suggest that GNz11 has a slightly harder radiation field at energies higher than the He$^+$ ionizing edge. As such the interpretation of GN-z11 based on these line ratios is more ambiguous, as it is broadly consistent with both AGN and SF model predictions \citep[see discussion in ][]{Bunker2023, Maiolino2023}.
However, the line ratios place RXCJ2248-ID in the star-formation regime of the CIII]/HeII vs. CIV/CIII] diagram presented by \citet{Scholtz2023} (based on models from \citet{Feltre2016} and \citet{Gutkin2016}), distinguishing it from AGN and from composite systems where the ionizing mechanism may be ambiguous. A more comprehensive comparison to various line ratio diagnostics will be provided by Plat et al. (2024), in prep. 

In addition to the line ratios described above, we seek 
emission lines that probe the ionizing spectrum at energies above 54 eV. 
The [Ne IV]$\lambda\lambda2422,2424$ emission line provides our 
best probe in the near-UV, constraining presence of the 63.5 eV 
photons required to triply ionize neon.
The models of \citet{Feltre2016} indicate that the $\log(\rm Ne\textsc{iv}/C\textsc{iv})$ ratio characteristic of AGN and SF differ significantly. 
The Ne IV line is undetected in the spectrum of RXCJ2248-ID at the expected location of 17210-17224\angstrom{}, allowing us to place a $3\sigma$ upper limit on the flux of $<1.3\times10^{-19}~\rm erg/s/cm^2$.
This constraint results in a 3$\sigma$ line ratio upper limit of $\log(\rm Ne\textsc{iv}/C\textsc{iv})<-1.6$.
This is inconsistent with the expectation from AGN models \citep{Feltre2016} which typically present values of $\log(\rm Ne\textsc{iv}/C\textsc{iv})\simeq-1.5$ to $0.2$. In contrast, the limit is consistent with the SF models which lie at values of $<-2.0$.

\citet{Scholtz2023} have recently demonstrated the diagnostic power of such high-ionization lines.
They demonstrated that star-forming and AGN photoionization models separate cleanly in the CIII]/HeII vs. [NeIV]/CIII] and CIII]/HeII vs. [NeV]/CIII] diagrams.
In these two diagnostics, our strict upper limits suggest RXCJ2248-ID is distinct from the AGN regime, and are consistent with photoionization from stars.
We will discuss this in more detail by Plat et al. (2024), in prep. Here we simply emphasize that the pattern of high-ionization lines in RXCJ2248-ID are not inconsistent 
with photoionization by metal poor massive stars.

\begin{table}
\caption{Rest-optical line observed-frame wavelengths, line fluxes, equivalent widths, and FWHM derived from fitting Gaussian profiles to emission lines following the procedure outlined in Section~\ref{sec:measurements}. FWHM have not been corrected for instrumental effects. Upper limits presented here are set at the $3\sigma$ level.
\newline $^{\rm a}$ The [OII]$\lambda\lambda3727,3729$ lines are blended and we cannot measure line widths individually.
\newline $^{\rm b}$ While [OIII]$\lambda6302$ is not formally detected at $>3\sigma$, we note the peak of the weak feature present at this wavelength.}
\begin{center}
\renewcommand{\arraystretch}{1.2}
\begin{tabular}{ccccc}
\toprule
Line & $\lambda_{\rm obs}$  & Flux  & EW  & FWHM \\
 &$\rm \angstrom{}$ & $\rm 10^{-19}erg/s/cm^2$ &\angstrom{} & km/s\\
\midrule
$\rm [OII]\rm \lambda\lambda3728$& 26488.4 & $1.1^{+0.4}_{-0.3}$ & $14.2^{+6.4}_{-4.1}$ & $-^{\rm a}$\\
$\rm [NeIII]\rm \lambda3869$&27495.6       & $19.8^{+0.4}_{-0.4}$ & $253^{+20 }_{ -16 }$ & $287^{+52 }_{ -61 }$\\
$\rm HeI\rm \lambda3890$&27644.0           &  $3.5^{+0.3}_{-0.3}$ & $44.1^{+6.2}_{-5.2}$ & $287^{+52 }_{ -61 }$\\
$\rm [NeIII]\rm \lambda3968$&28197.2       & $6.8^{+1.3}_{-0.9}$ & $87.2^{+18.9}_{-18.1}$ & $316^{+127 }_{ -94 }$\\
H$\epsilon$&28214.5                         &  $5.4^{+0.4}_{-0.4}$ & $68.7^{+16.2}_{-15.8}$ & $316^{+127 }_{ -94 }$\\
H$\delta$&29155.1                           &  $7.2^{+0.3}_{-0.3}$ & $91.9^{+14.2 }_{ -15.6 }$ & $344^{+78 }_{ -62 }$\\
H$\gamma$&30850.2                           & $13.0^{+0.2}_{-0.2}$ & $166^{+10 }_{ -10 }$ & $377^{+53 }_{ -43 }$\\
$\rm OIII\rm \lambda4364$&31014.4           &  $10.7^{+0.2}_{-0.2}$ & $137^{+9 }_{ -8 }$ & $377^{+53 }_{ -43 }$\\
$\rm HeI\rm \lambda4471$&31793.1             &  $2.4^{+0.2}_{-0.2}$ & $27.7^{+21.5}_{-12.2}$ & $403^{+60 }_{ -96 }$\\
$\rm HeII\rm \lambda4686$               & - &  $<0.9$ & $<8.2$ & $-$\\
H$\beta$&34549.9 & $25.4^{+0.2}_{-0.2}$ & $324^{+44 }_{ -35 }$ & $331^{+32 }_{ -47 }$\\
$\rm [OIII]\rm \lambda4960$&35243.8 &  $52.3^{+0.2}_{-0.2}$ & $668^{+90 }_{ -59 }$ & $321^{+37 }_{ -44 }$\\
$\rm [OIII]\rm \lambda5008$&35583.9 &  $166.9^{+0.2}_{-0.2}$ & $2130^{+70 }_{ -63 }$ & $318^{+53 }_{ -39 }$\\
$\rm HeI\rm \lambda5877$&41761.4 &  $5.2^{+0.2}_{-0.2}$ & $66.8^{+17.6 }_{ -19.6 }$ & $389^{+64 }_{ -94 }$
\\
$\rm [OI]\rm \lambda6302$&$44762.0^{\rm b}$ & $<2.2$ & $<27.4$ & $-$\\
H$\alpha$ &46646.0 &  $64.7^{+0.2}_{-0.2}$ & $826^{+41}_{-43}$ & $275^{+52 }_{ -49 }$\\
$\rm [NII]\rm \lambda6585$ &46788.5 &  $1.6^{+0.2}_{-0.2}$ & $20.1^{+4.7}_{-4.2}$ & $284^{+75 }_{ -105 }$\\
$\rm [SII]\lambda6717$  &- & $<0.7$ & $<13.3$ & $-$ \\
$\rm [SII]\lambda6730$  &- & $<0.7$ & $<13.3$ & $-$ \\
$\rm HeI\lambda7065$ &50204.6 & $3.1^{+0.5}_{-0.5}$ & $49.1^{+24.2}_{-19.1}$ & $223^{+33 }_{ -24 }$ \\
 \bottomrule
 \end{tabular}
 \end{center}

\label{tab:opticallines}
\end{table}

\subsection{Rest-frame optical emission lines}
\label{sec:optlines1}

The rest-frame optical contains a series of 
commonly-used nebular and auroral emission lines that characterize the properties of ionized gas \citep[e.g.,][]{Sanders2016, Kashino2017, Kewley2019, Sanders2023b, Strom2023}. Prior to {\it JWST}, our view of the $z>6$ universe was entirely limited to the rest-frame UV, stunting our interpretation of the high ionization line emission. Our G235M and G395M observations of RXCJ2248-ID deliver our first view of the rest-frame optical 
emission lines in this galaxy, probing rest-frame wavelength of $2336$ to $4460$\angstrom{} and $4039$ to $7310$\angstrom{}. We use these data to explore  the stars and gas conditions which are linked to such an intense 
radiation field. 
In this section, we present the rest-optical emission line measurements and discuss what they imply about the nature of RXCJ2248-ID.

\begin{figure*}
    \centering
     \includegraphics[width=1.0\linewidth]{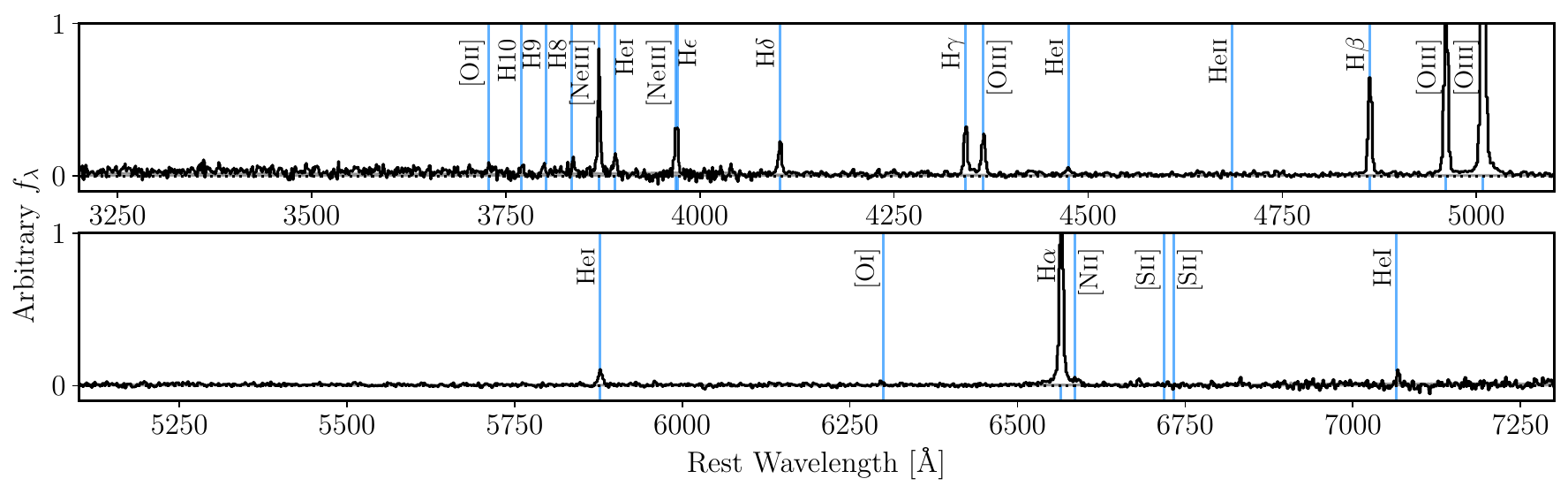}
     \caption{Rest-frame optical spectrum of RXCJ2248-ID. We display the error spectrum as a grey shaded region. All of the emission features detected at $>3\sigma$ are labeled and indicated by a blue vertical lines. We also mark the location of lines discussed in Section~\ref{sec:optlines1} that are not detected (e.g., HeII$\lambda4686$, [SII]$\lambda\lambda6717,6730$). Measurements for each of these lines are tabulated in Table~\ref{tab:opticallines}. This spectrum covers wavelengths in the rest-frame of  3200--7300\angstrom{}.} 
     \label{fig:optspectrum}
\end{figure*}

\subsubsection{Dust attenuation sensitive line ratios}
We constrain the dust attenuation in RXCJ2248-ID using the strongest hydrogen Balmer lines.
The H$\alpha$ and H$\beta$ lines are detected with extremely high S/N ($>100$) at observed-frame wavelengths of 46646.0\angstrom{} and 34549.9\angstrom{}, respectively.
Both lines fall within the G395M grating, and are thus immune from uncertainties due to inter-grating systematics.
Using the measured fluxes of these lines we derive a Balmer decrement of $\rm H\alpha/H\beta=2.55^{+0.06}_{-0.05}$.
This value is consistent with the absence of dust attenuation, but is slightly below the canonical value of $\rm H\alpha/H\beta=2.86$ \citep{Osterbrock} which assumes an electron temperature and density of $T_e=10$kK and $n_e=100~\rm cm^{-3}$, respectively.
However, the derived Balmer decrement is in better agreement with the value expected from Case B recombination theory \citep[e.g.,][]{Hummer1987} for the temperature and density we derive for RXCJ2248-ID ($T_e=24600K$, $n_e=10^5/\rm cm^3$; see \S\ref{sec:derived} below) of $\rm H\alpha/H\beta=2.70$.
We discuss the effects of density and temperature on the observed emission lines in detail in Sections~\ref{sec:abundances} and \ref{sec:density}, respectively.
We find a consistent result on the dust attenuation when considering higher-order Balmer lines (e.g., H$\gamma$, H$\delta$) however with larger uncertainties.

\subsubsection{[OIII] and H$\alpha$ emission line profiles}
\label{sec:o3ha}
The rest-optical spectrum of RXCJ2248-ID exhibits intense [OIII]$\lambda\lambda4959,5007$ and H$\alpha$ emission.
We measure a total [OIII]$\lambda5007$ ([OIII]$\lambda4959$) flux of $1.67\times10^{-17}~\rm erg/s/cm^2$ ($0.52\times10^{-17}~\rm erg/s/cm^2$), yielding an [OIII]$\lambda\lambda4959,5007$ EW of $2800^{+115}_{-87}$\angstrom{}. This approaches the highest EWs typically found in photoionization model grids of stellar populations \citep[$\simeq4500\angstrom$, e.g.,][]{Gutkin2016, Topping2022b}.  
Similarly, the prominent H$\alpha$ emission presents an integrated EW (flux) of $825.5\angstrom{}$ ($0.65\times10^{-17}~\rm erg/s/cm^2$).
This H$\alpha$ EW is slightly above the typical values seen at $z\sim6$ \citep[i.e., $780$\angstrom{};][]{Endsley2023b}, which is consistent with its presence significantly above the star-forming main sequence and young inferred age of its stellar population (see Section~\ref{sec:stars}).
Furthermore, the high [OIII]/H$\alpha$ is indicative of a hard ionizing radiation field \citep[e.g.,][]{Endsley2023b}. 
These line strengths suggest RXCJ2248-ID can be firmly classified as one of the most extreme emission-line galaxies (EELG) known, occupying a regime rarely seen at lower redshift \citep[e.g.,][]{Tang2019}, or in low-metallicity dwarf galaxies in the local Universe \citep[e.g.,][]{Flury2022a}.
Current reionization era samples include only a small population of galaxies with [OIII] EWs comparable to RXCJ2248-ID; such objects exist $>2\sigma$ above the median $\rm [OIII]+H\beta$ EW at $z\sim6$ \citep[$930\angstrom{}$ at $\rm M_{\rm UV}=-20$;][]{Endsley2023b}.
Thus, despite an increasing prevalence of EELGs in the galaxy population at early times, RXCJ2248-ID stands out as one of the strongest line emitters of this epoch.

\begin{figure*}
    \centering
     \includegraphics[width=1.0\linewidth]{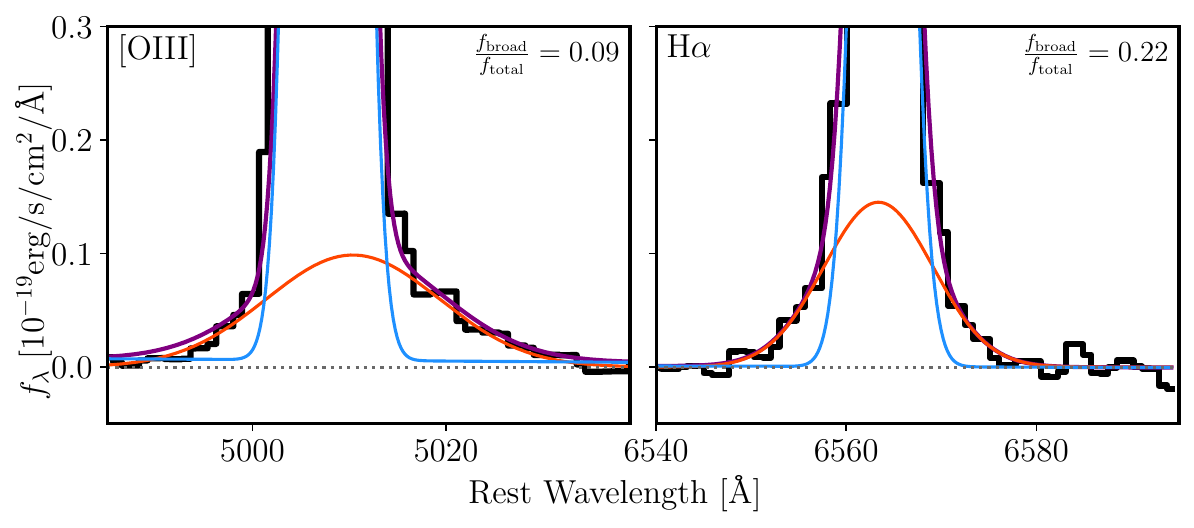}
     \caption{Decomposition of narrow and broad line fits to [OIII] (left) and H$\alpha$ (right) lines. The observed spectrum is displayed by the black histogram, while the red and blue lines represent the best-fit profiles to the broad and narrow components, respectively. The purple line shows the sum of the two component fits. These spectra are presented with a linear fit to the underlying continuum removed.  For each line, we provide the fraction of total flux contained within the broad component of the emission.} 
     \label{fig:optbroad}
\end{figure*}

A close inspection of the  H$\alpha$ profile reveals a broad component in emission.
In Figure~\ref{fig:optbroad} we decompose the observed H$\alpha$ line into broad and narrow components, and a summary of the component properties is presented in Table~\ref{tab:broad}.
We separate the broad component by fitting a Gaussian profile to the wings in the emission line. We then subtract this fit from the observed profile, leaving only the narrow emission component.
From our best-fit profiles, we find that the centroid of the narrow and broad components lie at a consistent redshift. 
The FWHM of the narrow component is $241^{+42}_{-38}~\rm km/s$, which is comparable to both the instrument resolution and other nebular lines in the spectrum.
However, the broad component is significantly wider, for which we measure a FWHM of $607^{+42}_{-36}~\rm km/s$.
In total, the broad component comprises $22\%$ of the overall H$\alpha$ flux.
This contribution from broad emission is consistent with the range of values found for local Green Pea galaxies \citep[e.g.,][]{Amorin2012, Llerena2023} and in high-redshift star-forming galaxies \citep[e.g.,][]{Davies2019, Freeman2019}. 
We note that we do not see broad H$\beta$ in our spectrum. However if we assume the H$\alpha$ and H$\beta$ lines present similar contributions from a broad component (i.e., $f_{\rm broad}/f_{\rm narrow}(\rm H\beta)=0.22$), we would not expect broad H$\beta$ to be detected given the S/N. 

The [OIII] emission also shows a broad component to its emission profile (Figure ~\ref{fig:optbroad}).  
Following our methodology described above for H$\alpha$, we fit the broad [OIII] emission wings to extract properties of the broad component.
Despite both [OIII] and H$\alpha$ showing broad emission components, their properties show several differences.
First, the broad [OIII] emission comprises only $8\%$ of the total [OIII] flux, in contrast to the 22\% measured for H$\alpha$.
Additionally, the broad [OIII] emission extends to  higher velocities than the broad H$\alpha$; we measure a FWHM for this component of $1258^{+60}_{-49}~\rm km/s$. This value is fully consistent with the upper bound of broad [OIII] line widths seen in star forming galaxies \citep{Davies2019,ForsterSchreiber2020,Tang2023,Carniani2023}.
Finally, we find an offset between the centroids of the broad and narrow [OIII] emission, such that the broad emission is redshifted by $157^{+30}_{-37}~\rm km/s$ relative to the narrow component.
This is similar to the velocity offset between broad and narrow [OIII] emission seen in other star forming galaxies (e.g., \citealt{Freeman2019, Hogarth2020, Tang2023}). While we cannot rule out the contribution of AGN energetics to the outflows, the broad line profiles appear consistent with expectations for outflows from star forming systems.

\subsubsection{Ionization-sensitive line ratios}
We combine high and low-ionization rest-optical emission lines to investigate the ionization state of the ISM.
Within our wavelength coverage, [NeIII] and [OIII] are the best indicators of high-ionization emission, while [OII]$\lambda\lambda3727,3729$ and [SII]$\lambda\lambda6717,6730$ probe the low ionization regime.
The [SII] lines are not detected ($3\sigma$ upper limit for each doublet member of $<0.7\times10^{-19}~\rm erg/s/cm^2$). We do detect very faint emission from [OII] with a total [OII] flux of $1.1\times10^{-19}~\rm erg/s/cm^2$ (S/N$=3.7$).
Section 4.2.2 demonstrates the presence of strong [OIII] emission; [NeIII]$\lambda3869$ follows as one of the strongest rest-optical lines in RXCJ2248-ID, with an EW of 252.7\angstrom{}.
In combination, these emission line measurements yield an extremely high O32 ($\rm \equiv[OIII]/[OII]$) of $184^{+88}_{-37}$, a Ne3O2 ($\rm \equiv[NeIII]/[OII]$) of $16.6^{+8.3}_{-3.3}$, and a low O2 ($\rm \equiv[OII]/H\beta$) of $0.04^{+0.02}_{-0.01}$.
These O32 and Ne3O2 line ratios are consistent with an ISM under extremely high ionization conditions.
High gas densities can mimic these conditions, as the low ionization (e.g., [OII]) lines are suppressed by collisional de-excitation.
We return to discuss the impact of the ISM density in Section~\ref{sec:density}.

\begin{figure*}
    \centering
     \includegraphics[width=1.0\linewidth]{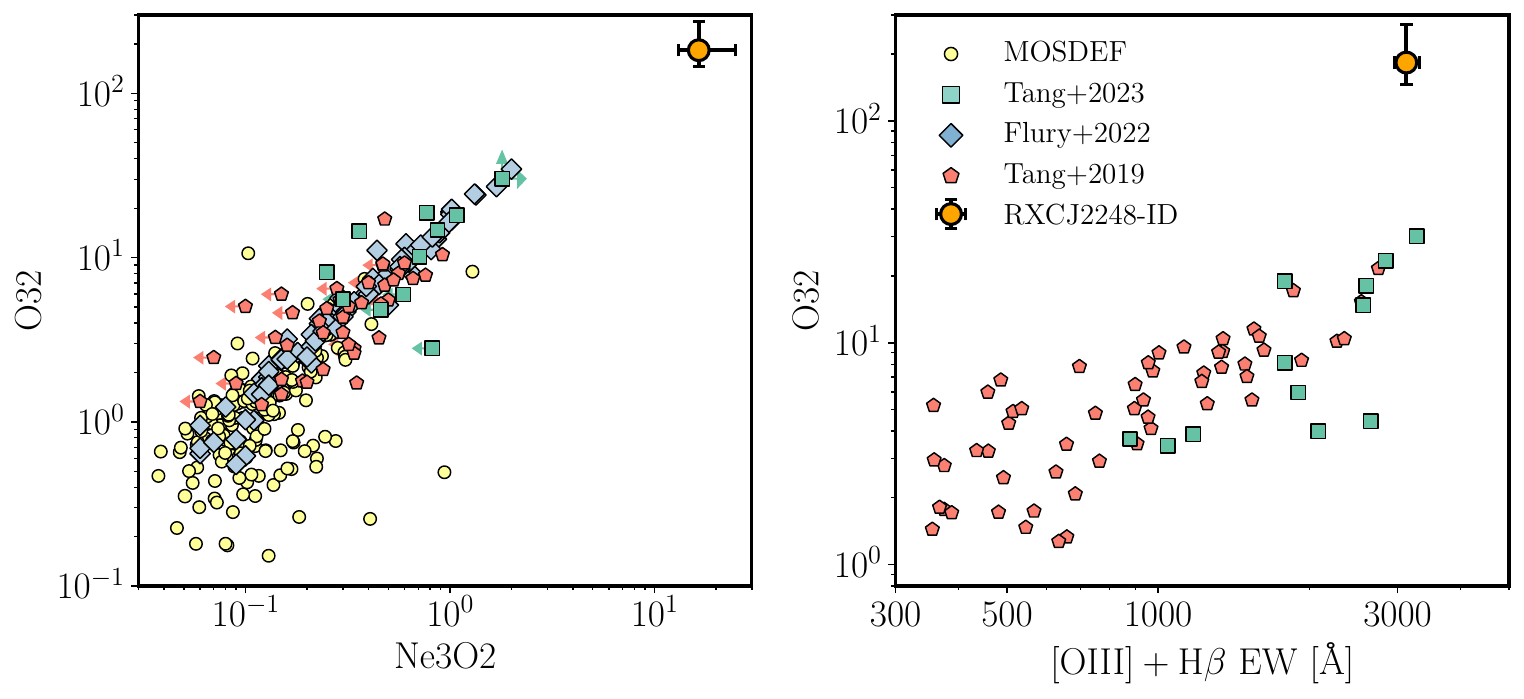}
     \caption{Line ratios and equivalent widths of RXCJ2248-ID (orange circle) compared to literature sources in the local Universe and at high redshift. O32 versus Ne3O2 is shown in the left panel, while O32 versus $\rm [OIII]+H\beta$ EW is displayed in the right panel.  We additionally show comparison samples derived from the literature. Measurements at in the local Universe from the LzLCS \citep{Flury2022a} are shown as the blue diamonds. Galaxies at $z\sim1-3$ from the MOSDEF survey \citep{Kriek2015, Jeong2020}, and from \citet{Tang2019}, are presented as yellow circles and red pentagons, respectively. Finally, we compare to spectroscopic measurements of galaxies at $z\sim7-9$ from \citet{Tang2023} denoted by the green squares.} 
     \label{fig:o32}
\end{figure*}

These line ratios are drastically offset from values typically seen in star-forming galaxies.
Figure~\ref{fig:o32}a compares the O32 and Ne3O2 ratios of RXCJ2248-ID to samples of galaxies across cosmic time.
The O32 value that we measure for RXCJ2248-ID ($\rm O32=184^{+88}_{-37}$) is roughly two orders of magnitude above typical values observed at $z\sim2-3$ \citep[$\rm O32=1-2$; e.g.,][]{Steidel2014, Sanders2016}, and more than a factor of $20\times$ greater than the most intense line emitters \citep[e.g.,][]{Maseda2018, Tang2019}.
Despite the increase in average O32 at $z\gtrsim6$ \citep[$\rm O32=18$][]{Tang2023, Mascia2023, Sanders2023, Cameron2023} relative to $z\sim2$, the value for RXCJ2248-ID remains $10\times$ above objects in the reionization era. 
The Ne3O2 ratio follows an almost identical trend, such that RXCJ2248-ID is $110-170\times$ and 19$\times$ times larger than typical values found at $z\sim2$ (Ne3O2$\sim0.10-0.15$, \citealt{Steidel2016, Sanders2021}) and $z\gtrsim6$ (Ne3O2$\sim0.9$, \citealt{Tang2023, Cameron2023, Sanders2023}), respectively.

RXCJ2248-ID is clearly characterized by high O32 and Ne3O2, as well as extremely large $\rm [OIII]+H\beta$ EWs.
These two trends are known to form a tight correlation \citep{Tang2019, Sanders2020}, however RXCJ2248-ID is seen as a significant outlier in the latter relation.
We display O32 as a function of $\rm [OIII]+H\beta$ EW in Figure~\ref{fig:o32}b.
At the $\rm [OIII]+H\beta$ EW of RXCJ2248-ID, few galaxies exist (see \S~\ref{sec:o3ha}), however the few known examples display O32 ratios in the range 3-30 (median of 18).  This is a factor of $\times10$ below our measured value of O32=$184^{+88}_{-37}$.  This may provide further indication that the O32 ratio in RXCJ2248-ID  is elevated as a result of processes beyond the ionization conditions (e.g., density; see \S~\ref{sec:density} below).

While noting the potential influence of the high electron density on the [OII] flux, 
the large O32 ratio nonetheless likely points to a large ionization parameter ($\log(U)$). 
As a first step, we consider models with typical electron densities ($100$ cm$^{-3}$) calibrated by \citet{Berg2019}, and assume a metallicity of $1/20Z_{\odot}$.
The metallicity is in good agreement with the values inferred for RXCJ2248-ID (see Section~\ref{sec:abundances} below).
For the measured O32 of $184$, we derive an ionization parameter of $\log(U)=-1.0$ based on the \citet{Berg2019} calibrations.
This derivation is relatively insensitive to the metallicity that we assume; introducing a factor of two difference in the metallicity results in a change in ionization parameter of just $\Delta \log(U)=0.2$.
As alluded to above, and discussed in detail in \S\ref{sec:density}, the ISM density may be contributing to the extremely high O32 value. However photoionization models that consider variations in electron density still find large ionization parameters ($\log(U)=-0.5$) are required to reproduce the emission lines (Plat et al. 2024, in prep). 
This intense ionization parameter is likely driven by the high density of UV photons in the compact region 
targeted by spectroscopy and will play a critical role in setting overall abundances in Section~\ref{sec:abundances}.

\subsubsection{[OIII]$\lambda$4363 auroral line}

\jwst{}/NIRSpec has proven to be effective at detecting the weak [OIII]$\lambda4363$ line at $z>6$ \citep[e.g.,][]{Sanders2023, Curti2023}.
We detect this line at high significance (S/N$=21$) in our spectrum, and measure flux of $10.7^{+0.2}_{-0.2}\times10^{-19}~\rm erg/s/cm^2$.
This line is comparable to the strength of the nearby H$\gamma$ line, with a flux ratio of $f_{\rm [OIII]\lambda4363}/f_{\rm H\gamma}=0.82^{+0.03}_{-0.04}$, which is indicative of high electron temperatures in the ISM.
Among the few examples of $z>6$ star-forming galaxies with [OIII]$\lambda4363$ of comparable strength to RXCJ2248-ID ($\rm EW=137^{+9}_{-8}\angstrom{}$), temperatures in excess of $20$kK are implied \citep[e.g.,][]{Sanders2023, Curti2023}.
In Section~\ref{sec:abundances}, we provide a detailed calculation of the electron temperature and implied abundances using these line measurements.

\subsubsection{He I and He II emission lines}

We detect a series of He I recombination lines at $>3\sigma$ with rest wavelengths of 3889\angstrom{}, 4471\angstrom{}, 5876\angstrom{}, and 7065\angstrom{}.
The strength of these lines relative to H$\beta$ (i.e., $f_{3889}/f_{\rm H\beta}=0.14$, $f_{4471}/f_{\rm H\beta}=0.09$, $f_{5876}/f_{\rm H\beta}=0.21$, $f_{7065}/f_{\rm H\beta}=0.12$) is far greater than can be explained by the relative He/H abundance from photoionization alone \citep[e.g.,][]{Izotov1998}.
In addition, these flux fractions are a factor of $2-4\times$ greater than what is observed in nearby EELGs \citep{Berg2021},
further supporting the presence of an additional mechanism driving their emission.
The most important mechanism that can further boost the strength of these lines is collisional excitation \citep{Clegg1987}.
The observed He I strengths may further support the high temperatures and densities indicated by the rest-UV and rest-optical line measurements in RXCJ2248-ID.

In spite of the inferred hard radiation field, we do not detect the He II$\lambda4686$ 
recombination line in the rest-frame optical. 
The spectrum of RXCJ2248-ID yields a $3\sigma$ upper limit of $<0.9\times10^{-19}~\rm erg/s/cm^2$, corresponding to an EW limit of $<~8.2~$\angstrom{}.
Based on the observed flux of HeII$\lambda1640$ and assuming an intrinsic ratio (i.e., HeII$\lambda1640/\rm HeII\lambda4686=6.96$, \citealt{Dopita2003}), we expect a total flux of $f_{\rm HeII4686}=0.8\times10^{-19}~\rm erg/s/cm^2$, which is consistent with the $3\sigma$ flux upper limit that we derive.
The HeII$\lambda4686$ upper limit implies a flux relative to H$\beta$ that is less than $<0.035$ at the $3\sigma$ level.
Using the inferred HeII$\lambda4686$ flux based on the HeII$\lambda1640$ measurement, we would expect a flux ratio of $f_{\rm HeII4686}/f_{\rm H\beta}=0.03$.
Constraints from low-metallicity compact dwarf galaxies in the local Universe typically report fractions of $\simeq0.02$ \citep[e.g.,][]{Senchyna2017, Berg2021, Izotov2024}, and only rarely reaching above our upper limit.

\section{Gas-phase Properties of a \texorpdfstring{$z=6.1$}{z=6.1} CIV emitter}
\label{sec:derived}
In this section we investigate the gas-phase properties of RXCJ2248-ID that are implied by our measurements.
We use multiple density-sensitive line ratios to derive the electron density in Section~\ref{sec:density}, and discuss abundance patterns set by this spectrum in Section~\ref{sec:abundances}.

\begin{table}
\caption{Derived properties for RXCJ2248-ID. Stellar mass and SFR is derived using synthetic broadband photometric SED model fits, while the other values are inferred from the spectroscopic measurements.
}
\begin{center}
\renewcommand{\arraystretch}{1.2}
\begin{tabular}{cc}
\toprule
Quantity & Value  \\
\midrule
$\log(\rm M/M_{\odot})$ & $8.05^{+0.17}_{-0.15}$  \\
$\log(\rm SFR/M_{\odot}yr^{-1})$ & $1.8^{+0.2}_{-0.2}$  \\
$t_{\rm age}^{\rm CSFH}/\rm Myr$ & $1.8^{+0.7}_{-0.4}$\\
$\beta$ & $-2.72\pm0.16$  \\
$\tau_{\rm V}$ & $0.02^{+0.03}_{-0.02}$ \\
$\rm EW_{\rm [OIII]+H\beta}^{\rm SED}/\angstrom{}$&$ 3706^{+378}_{-505}$\\
$\log(\rm \xi_{\rm ion}/erg^{-1}~Hz)$&$25.94^{+0.11}_{-0.07}$\\
$\log(\rm U)$ & $-1.0\pm0.2$\\
$n_e(\rm Si\textsc{iii}])/cm^{-3}$ & $6.4^{+5.3}_{-2.6}\times10^4$  \\
$n_e(\rm C\textsc{iii}])/cm^{-3}$ & $1.1^{+0.1}_{-0.2}\times10^5$  \\
$n_e(\rm N\textsc{iv}])/cm^{-3}$ & $3.1^{+0.5}_{-0.4}\times10^5$  \\
$\rm T_e(\rm [OIII])/K$ & $24600\pm2600$  \\
$12+\log(\rm O/H)$ & $7.43^{+0.17}_{-0.09}$  \\
$\log(\rm C/O)$ &$ -0.83^{+0.11}_{-0.10}$  \\
$\rm [C/O]$ & $-0.60^{+0.11}_{-0.10}$\\
$\log(\rm N/O)$ & $-0.39^{+0.10}_{-0.08}$  \\
$\rm [N/O]$ & $0.47^{+0.10}_{-0.08}$\\
\bottomrule
\end{tabular}
\end{center}

\label{tab:properties}
\end{table}

\subsection{Electron density of high ionization gas}
\label{sec:density}

The spectrum of RXCJ2248-ID contains multiple density-sensitive line ratios.
Each of these emission lines probes distinct ionization states of the ISM \citep[see e.g.,][]{Kewley2019}.
The density of low-ionization gas is typically constrained by the [OII] and [SII] doublets.  
However we are unable to use these lines to estimate densities, as [SII] is not detected and [OII] cannot be deblended at the resolution of spectrum.
Density-sensitive line ratios that can be utilized are Si III], C III], and N IV], which have ionization energies of 16.3 eV, 24.4 eV, and 47.4 eV, respectively.
We now explore the electron densities inferred from each of these line ratios.

\begin{figure*}
    \centering
     \includegraphics[width=1.0\linewidth]{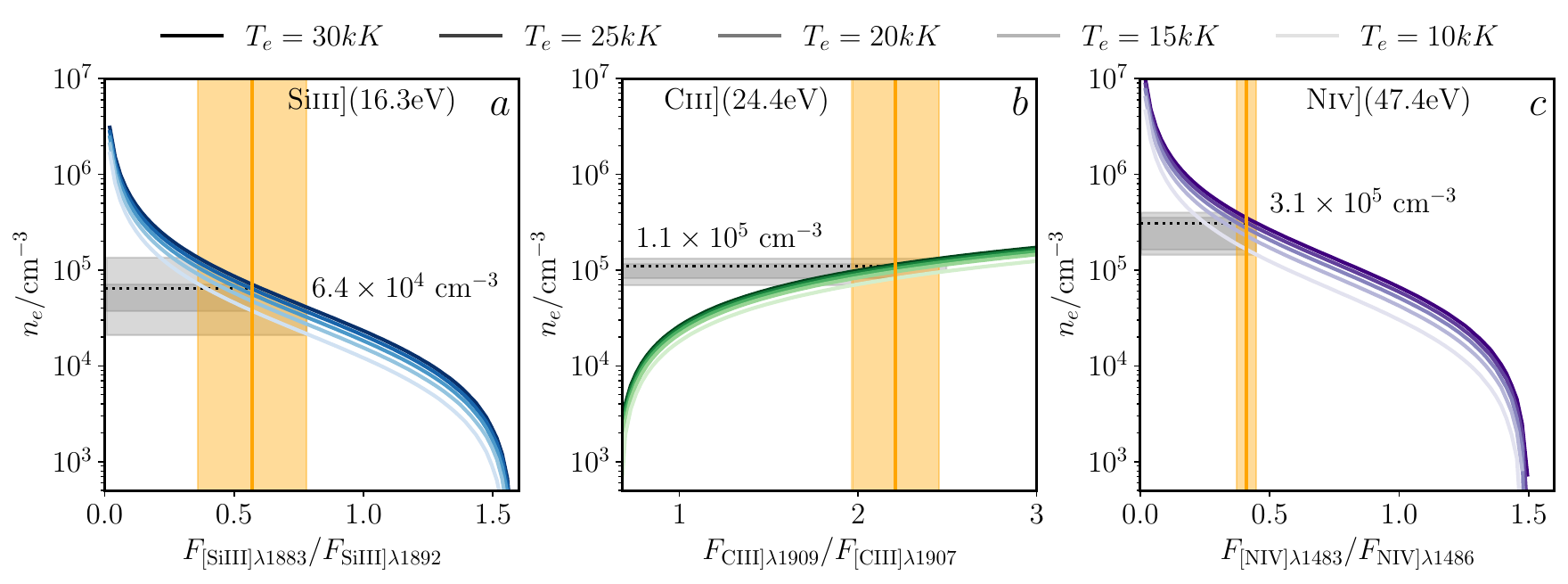}
     \caption{Electron Densities inferred from the density-sensitive SiIII] (panel a),  CIII] (panel b), and NIV] (panel c) line ratios. The panels are ordered by ionization energy, which is noted at the top of each plot. The colored curves are shaded according to the electron temperature that was assumed ($10-30$kK), which are indicated at the top of the figure. The measured line ratios and corresponding uncertainties are indicated by the vertical orange line and orange shaded regions, respectively. The horizontal black dotted line indicates the density derived from each line ratio, while the dark and light grey shaded regions display the $1\sigma$ and $2\sigma$ confidence intervals, respectively.} 
     \label{fig:density}
\end{figure*}

We begin by deriving electron densities using the SiIII]$\lambda1883$/SiIII]$\lambda1892$ line ratio \citep[e.g.,][]{Dufton1984}.
This line ratio is a useful density indicator for densities below $n_e \lesssim 10^{11}~\rm cm^{-3}$ and above $n_e \gtrsim 10^{3}~\rm cm^{-3}$.
In Figure~\ref{fig:density}a we display the electron density that we recover as a function of measured line ratio.
We repeat this calculation for different electron temperatures in the range 10-30kK.
While in the following section (\S~\ref{sec:abundances}) we find that the electron temperature is consistent with 25kK, the value that we assume when deriving densities has little effect over this range.
We detect both SiIII] components at S/N$>3$ (see Section~\ref{sec:uvlines}), and calculate a line ratio of $f_{1883}/f_{1892} = 0.57\pm0.21$.
At an electron temperature of 25kK, this line ratio corresponds to a density of $n_e=(6.3^{+5.3}_{-2.6})\times10^4~\rm cm^{-3}$.
Varying the electron temperature over the range of 10-30kK yields densities of $n_e=(3.7-6.9)\times10^4~\rm cm^{-3}$, which is less variation than that from the measurement uncertainty.

Emission from CIII] constrains the ionizing spectrum at higher energies (24.4 eV) compared to that implied by SiIII] (16.3 eV).
As such, we expect the densities implied by CIII] to be greater than that of SiIII].
We display the electron density derived as a function of measured line ratio in Figure~\ref{fig:density}b.
From these calculations, we see that ratios of CIII]$\lambda1909$/[CIII]$\lambda1907>1$ imply densities greater than $n_e \gtrsim 2\times10^{4}~\rm cm^{-3}$, with large variations in line ratio leading to small changes in density.
We measure a line flux ratio of CIII]$\lambda1909$/[CIII]$\lambda1907=2.21\pm0.11$.
From this line ratio, we infer a density of $n_e=(1.1^{+0.1}_{-0.2})\times10^5~\rm cm^{-3}$ (when assuming $T_e=25$ kK).
Over the full range of considered temperatures (10-30kK) we find densities ranging from $n_e=(0.83-1.15)\times10^5~\rm cm^{-3}$.

The final density-sensitive line ratio that we explore is the NIV] doublet.
Recent efforts to constrain the NIV] doublet in GN-z11 have only yielded a detection of NIV]$\lambda1486$, and provided an upper limit on the NIV]$\lambda1483$ component.
The line ratio implied by the resulting constraints ($F_{1483}/F_{1486}<0.48$ at $3\sigma$) is in the high density regime, leading to an electron density in GN-z11 above $>10^5~\rm cm^{-3}$ \citep{Senchyna2023, Maiolino2023}.
We detect both components in RXCJ2248-ID, and measure an emission-line ratio of $f_{1483}/f_{1486}=0.41\pm0.09$ (see Section~\ref{sec:uvlines}).
We show how this ratio corresponds to an inferred density in Figure~\ref{fig:density}c for a given temperature.
For our fiducial temperature of 25kK, this results in a density of $n_e=3.1^{+0.5}_{-0.4}\times10^5\rm cm^{-3}$.
This density calculation is relatively insensitive to the assumed temperature at $>15$ kK, where the values range from $n_e=2.3-3.5\times10^5\rm cm^{-3}$. Assuming a temperature of just 10kK results in a somewhat lower density of $n_e=1.6\times10^5\rm cm^{-3}$.
This result completes the trend between electron density and ionization state of the ISM observed within RXCJ2248.

The high densities in the gas-phase have widespread effects on the emission from RXCJ2248-ID.
Previous analyses have demonstrated that densities inferred from C III] are moderately high, and can range from $5\times10^3-25\times10^3~\rm cm^{-3}$ \citep[e.g.,][]{James2014, James2018}.
However, RXCJ2248-ID is nonetheless distinct in that it sees a shift not only to a higher C III] density ($110\times10^3~\rm cm^{-3}$), but also that its line emission is almost entirely resulting from this high-density, high-ionization regime.
As a consequence, we see significantly suppressed emission from low-ionization lines (e.g., [OII]), while boosting emission from others (e.g., He I) from collisional excitation.
These effects can in turn impact measured (e.g., O32) and derived (e.g., $\log(U)$) quantities, which are critical to our understanding of the conditions within these systems.
Further progress toward extending these calculations to regions of lower ionization and density with lines such as [OII], or higher density (with e.g., [NeIV] or [ArIV], \citealt{Kewley2019}) will require deeper spectroscopy.
It is clear however, that these lower and higher density ISM phases have a minimal contribution to the nebular emission from RXCJ2248-ID.

\subsection{Oxygen, carbon, and nitrogen abundance}
\label{sec:abundances}

From the emission line constraints provided in Sections~\ref{sec:uvlines} and \ref{sec:optlines1}, we derive carbon, nitrogen, and oxygen abundances.
We note that a full analysis of these abundances using photoionization models is presented by Plat et al. (in prep). 
We infer direct temperature oxygen abundances following the methodology presented in \citet{Izotov2006}.
First, we use the \texttt{pyneb} \citep{Luridiana2015} package to infer the electron temperature using the observed [OIII]$\lambda4363/\rm [OIII]\lambda5008$ flux ratio.
For this calculation we assume an electron density of $10^5~\rm cm^{-3}$, which is consistent with our results from Section~\ref{sec:density}.
As we see below, this assumed density has little impact in the resulting abundance determinations.
For this density, and a [OIII]$\lambda4363/\rm [OIII]\lambda5008$ ratio of $0.064$, we infer an electron temperatures of $24600\pm~2600\rm K$.
The electron temperature inferred using the rest-UV auroral OIII]$\lambda\lambda1660,1666$ ($T_e=23300\pm4100$ K) is consistent with this value, but has larger uncertainty.
This is among the hottest temperatures derived for star-forming galaxies at $z>6$, \citep{Curti2023, Schaerer2022, Sanders2023}, and very metal poor galaxies in the local Universe \citep{Izotov2018}.
However, these temperatures are consistent with being driven by star formation.
Based on these electron temperatures, densities, and [OIII]$\lambda5007/\rm H\beta$ flux ratios, we find an abundance of oxygen in the doubly-ionized state of $12+\log(\rm O^{++}/H)=7.43^{+0.17}_{-0.09}$.
If we assume a much lower density ($n_e=10^3~\rm cm^{-3}$) we find a change in $\rm O^{++}/H$ of only $0.03$ dex.

While we expect the $\rm O^{++}$ state to dominate the total oxygen abundance, we quantify the $\rm O^+$ and $\rm O^{3+}$ as well.
The extremely weak [OII] in RXCJ2248-ID suggests little contribution from the $\rm O^+$ zone.
While we do not have the measurements required to derive the [OII] temperature directly (e.g., using [OII]$\lambda\lambda7322,7332$), we follow \citet{Campbell1986} to infer it as $T_e(\rm [O\textsc{ii}])=0.7\times T_e(\rm [O\textsc{iii}])+3000K$.
For this $\rm O^+$ abundance calculation, we assume an electron density typical of the low-ionization zone ($n_e=300~\rm cm^{-3}$; \citealt{Isobe2023, Reddy2023, Sanders2023b}), and discuss how this assumption impacts the resulting abundance below.
These temperature and density assumptions yield an abundance of $12+\log(\rm O^+/H)=6.91$ in RXCJ2248-ID.
This $\rm O^+/H$ abundance is less than 10\% of the value found for $\rm O^{++}$/H (i.e., $12+\log(\rm O^+/H)\lesssim6$), confirming the minimal contribution of oxygen in this state.
Finally, we expect oxygen in the higher ionization states to comprise a very small fraction of the overall abundance, even in very intense ionization conditions \citep{Berg2018}.

We use the C\textsc{iv} and C\textsc{iii}] emission lines to infer the C/O abundance ratio following methods presented in the literature \citep[e.g.,][]{Berg2019, Jones2023}.
The carbon in these ionization states (i.e., C$^{++}$, C$^{3+}$) are expected to host the majority of the overall abundance.
The absence of very high ionization lines (e.g., [Ne IV]$\lambda\lambda2422,2424$) suggest that the contribution from C$^{\ge 4+}$ is minimal.
Constraining the C$^{3+}$ abundance directly from the CIV$\lambda1548$ flux may be subject to significant uncertainties, as the resonant nature of the line impeded accurate measurement of the total flux.
We therefore infer the total C/O ratio from the C$^{++}$/O$^{++}$ ratio to which we apply an ionization correction factor \citep[ICF; e.g.,][]{Berg2019}.
For this calculation, we fix the electron temperature and density (derived from CIII]) to values derived above (see Table~\ref{tab:properties}).
Using these assumptions and our measured $\rm CIII]\lambda\lambda1907,1909/OIII]\lambda\lambda1660,1666$ line ratio, we find an abundance of $\log(\rm C^{++}/O^{++})=-1.15^{+0.08}_{-0.07}$.
Using the calibrations provided by \citet{Berg2019}, we find that a metallicity of $1/20Z_{\odot}$ and $\log(U)=-1$ results in an ICF of $2.09$.
This in turn results in a sub-solar total $\log(\rm C/O)$ ratio of $-0.83^{+0.11}_{-0.10}$ ($\rm [C/O]=-0.60$).
As a comparison, we infer C/O abundances from the empirical calibration derived by \citet{PerezMontero2017} based on the C3O3 ($\equiv \log(\frac{f_{\rm CIII]\lambda1908}+f_{\rm CIV\lambda1549}}{f_{\rm OIII]\lambda1660}})$) line ratio. 
We derive a C3O3 ratio in RXCJ2248-ID of $0.27^{+0.06}_{-0.07}$, which corresponds to a relative abundance of $\log(\rm C/O)=-0.85$ ($[\rm C/O]=-0.62$). This is in excellent agreement with the value derived above ($\log(\rm C/O)=-0.83$), implying that this calibration may be appropriate even at high redshift.

Finally, we explore the N/O abundance ratio using the rest-UV N\textsc{iv}] and N\textsc{iii}] detections.
Due to the similarities in ionization energies of $\rm C^{++}$ and $\rm N^{++}$, it is straightforward to first derive a C/N abundance, and then use the C/O ratio to infer N/O \citep[see e.g.,][]{Berg2018}.
We derive a $\rm C^{++}/N^{++}$ ratio based on the $\rm NIII]\lambda1750$ and $\rm CIII]\lambda1908$ lines, and setting the temperature and density to that derived previously.
We note that variations in temperature and density to the extent of our uncertainties have only a minor effect on the inferred abundances.
The corresponding emission-line fluxes (Table~\ref{tab:UVlines}) result in a relative abundance of $\log(\rm C^{++}/N^{++})=-0.25$. We combine this inferred value with C/O to obtain a total $\log(\rm N/O)=-0.58^{+0.12}_{-0.10}$, exceeding the solar value by $0.28$ dex.
As an alternative estimate, we assume that the oxygen abundance is dominated by the $\rm O^{++}$ state, and that the total N/O abundance can be derived as:
\begin{equation}
    \rm \frac{N}{O} \simeq \frac{N^{++}+N^{3+}}{O^{++}}.
\end{equation}
We perform this calculation using the fluxes of $\rm NIII]\lambda1750$, $\rm NIV]\lambda\lambda1483,1486$, and $\rm OIII]\lambda\lambda1660,1666$, and find a super-solar abundance ratio of $\log(\rm N/O)=-0.39^{+0.10}_{-0.08}$ ($[\rm N/O]=0.47$).
This value is higher than our derivation that utilized the C/O ratio, however the results are broadly consistent to the level of uncertainty of our measurements.
In the former calculation, the relative $\rm N^{3+}/N^{++}$ abundance is implicitly inferred from from the C/O ICF, it is clear that the $\rm NIV]\lambda1483$ line is far stronger than $\rm NIII]\lambda1750$, and the assumed ICF may be an underestimate.

\begin{figure*}
    \centering
     \includegraphics[width=1.0\linewidth]{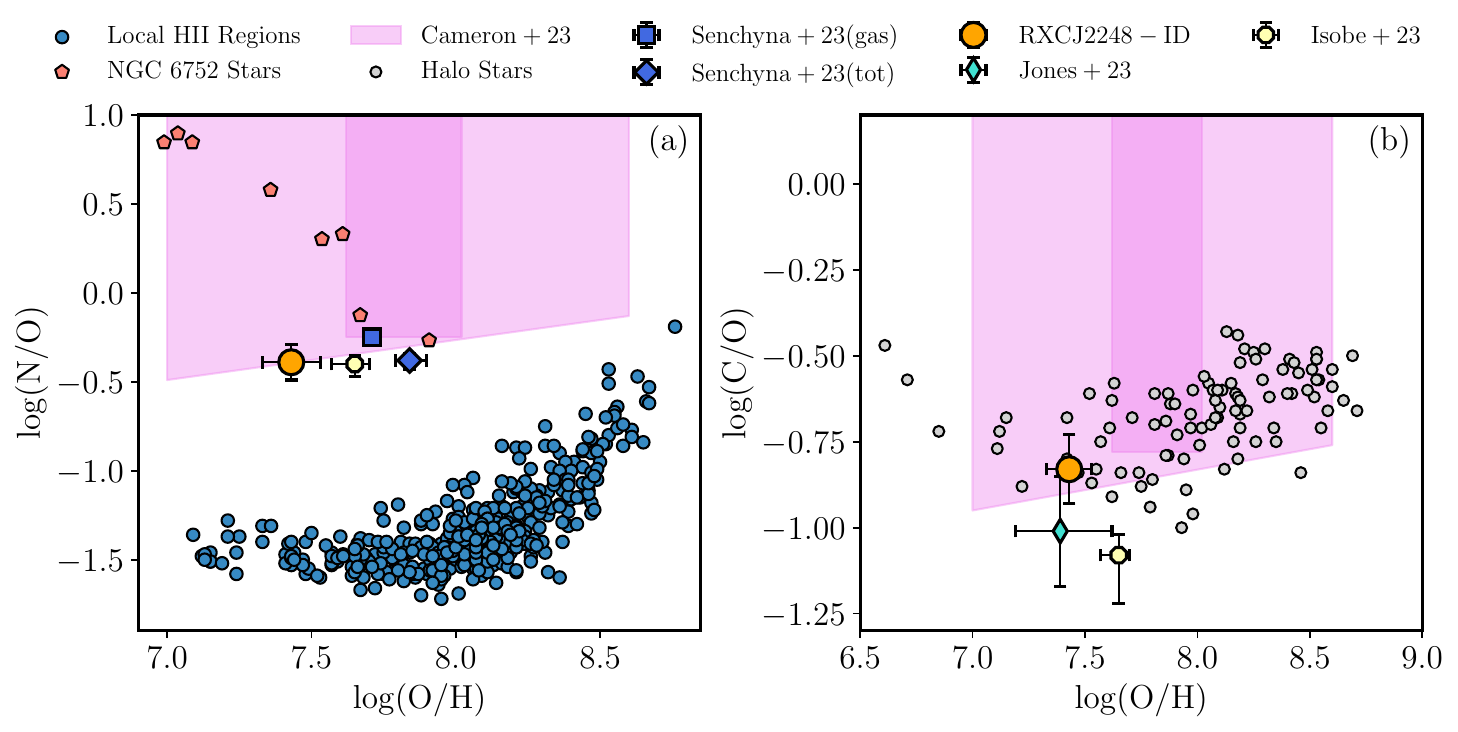}
     \caption{Comparison of abundance patterns of RXCJ2248-ID to other objects in the local Universe and at high redshift. Panel (a) displays the N/O abundance ratio as a function of O/H described in Section~\ref{sec:abundances}. The values derived for RXCJ2248-ID are displayed as an orange circle. Inferred total (gas-phase) abundances for GN-z11 derived by \citet{Senchyna2023} are displayed as the blue diamond (square). The fiducial (conservative) ranges of N/O and O/H inferred for GN-z11 from \citet{Cameron2023} is displayed as the light (dark) purple shaded region. Values for GLASS\_150008 at $z=6.23$ derived by \citet{Isobe2023} is shown as the yellow circle. The small blue circles represent abundances from local HII regions from \citet{Pilyugin2012}, and the red pentagons are measurements from stars in the globular cluster NGC 6752 \citep{Carretta2005}. RXCJ2248-ID has significantly higher nitrogen abundance at fixed O/H relative to the local sequence of HII regions, and is much more consistent with values probing conditions of globular cluster formation. Panel (b) compares C/O vs. O/H between RXCJ2248-ID and objects from the literature. We additionally compare to the C/O and O/H of GLASS\_150008 derived by \citet{Jones2023}. } 
     \label{fig:abundances}
\end{figure*}

The full set of abundances of RXCJ2248-ID is distinct from what is observed in typical systems (Figure~\ref{fig:abundances}).
At oxygen abundance of our galaxy, both RXCJ2248-ID and local HII regions inhabit a similar range of C/O ratios ($\log(\rm C/O)\simeq-0.6$ to $-1.0$).
The nitrogen abundance paints a different picture.
In low-metallicity HII regions ($12+\log(\rm O/H)<8.2$), primary nitrogen production is responsible for setting the N/O ratio \citep{Henry2000, Pilyugin2012} which is typically $\log(\rm N/O)=-1.5$ for all O/H.
RXCJ2248-ID is significantly enriched beyond this value, with an N/O that exceeds the low-metallicity plateau by 1 dex.
This arrangement of nominal carbon enrichment combined with an excess of nitrogen has recently been demonstrated for GN-z11 \citep{Senchyna2023, Cameron2023b} at $z=10.6$, and claimed based on lower S/N measurements for GLASS\_150008 and CEERS-1019 \citep{Isobe2023} at $z=6.2$ and $z=8.67$, respectively.
While distinct from local H II regions, similar and even larger N/O and N/C values are a well-established feature in the second population stars of globular clusters \citep[e.g.][]{Bastian2018}. We will discuss this in more detail in the next section.

\begin{table}
\caption{Measurements of the broad and narrow components of strong rest-optical emission lines in the spectrum.}
\begin{center}
\renewcommand{\arraystretch}{1.2}
\begin{tabular}{ccccc}
\toprule
Line & $\frac{f_{\rm Broad}}{f_{\rm Tot}}$ & FWHM$_{\rm narrow}$  & FWHM$_{\rm broad}$ & $\Delta v$\\
     &                             & [km/s]         & [km/s] & [km/s] \\
\midrule
$\rm H\alpha$ & $0.22_{-0.02}^{+0.02}$ & $242_{-29}^{+31}$ & $607_{-36}^{+42}$ & $-27_{-21}^{+23}$\\
$\rm [OIII]\lambda5007$ & $0.09_{-0.01}^{+0.01}$ & $311_{-30}^{+42}$ & $1258_{-49}^{+60}$ & $157_{-37}^{+30}$\\

\bottomrule
\end{tabular}
\end{center}

\label{tab:broad}
\end{table}

%
%
%
%
\section{Discussion}
\label{sec:disc}

A decade ago, ground-based telescopes provided our first view of reionization-era spectra, revealing strong emission from highly ionized species in the ISM   that are rarely seen at lower redshifts \citep[e.g.,][]{Stark2015a, Stark2015b, Laporte2017, Stark2017}. 
{\it JWST} has since built on these studies, revealing that hard radiation 
fields  accompany a significant subset of early star forming galaxies \citep{Tang2023, Bunker2023, Hsiao2023, Cameron2023, Mascia2023}. What causes reionization era galaxies to undergo this hard spectral phase has long remained unclear owing to the limitations of ground-based spectroscopy at $
z\gtrsim 6$.
In this paper, we have presented a comprehensive rest-UV to optical {\it JWST} investigation of RXCJ2248-ID, a multiply-imaged reionization era galaxy with strong CIV emission, one of two such systems known prior to {\it JWST}. As the galaxy is extremely bright owing to its magnification, the {\it JWST} data of RXCJ2248-ID provide our best template of the gas conditions and ionizing sources linked to the hard radiation field phase. 

The new results indicate that  RXCJ2248-ID is composed of two unresolved $\lesssim \rm{22}$ pc star forming complexes separated by $\simeq \rm{220}$ pc in the source plane. The CIV emission is powered by the brighter of the two clumps. The SED suggests an sSFR (560 Gyr$^{-1}$) which is comparable to GNz11 \citep{Bunker2023} and in the upper envelope of that found in $z\simeq 6-9$ galaxy samples \citep[e.g.,][]{Endsley2023b}. The compact CIV-emitting source in RXCJ2248-ID is host to a dense concentration of stars
(1.1$\times$10$^{8}$ M$_\odot$) and is undergoing vigorous star formation (63 M$_\odot$ yr$^{-1}$).  The stellar population is inferred to be very young (1.8 Myr), and the associated star 
formation rate surface density (10,400 M$_\odot$/yr/kpc$^2$) is a rarity at high redshift, more than 10$\times$ greater than the lower limit inferred for GNz11 \citep{Tacchella2023}. In addition, the high density of stellar mass in RXCJ2248-ID is in a regime reminiscent of individual star-forming clumps at high redshift (see Figure~\ref{fig:sizes}b). The star formation conditions in RXCJ2248-ID are conceivably a short-lived phase that many reionization-era galaxies go through as they experience a strong burst of star formation.

Diagnostic nebular lines in RXCJ2248-ID give direct insight into the 
gas properties which are linked to strong CIV emission.  
The gas is very metal poor ($12+\log(\rm O/H)=7.43^{+0.17}_{-0.09}$), boosting 
the strength of collisionally-excited lines. We measure high electron densities ($6.4-31\times10^4$ cm$^{-3}$) in three different transitions, perhaps a direct result of the high gas densities required to power such a strong burst of star formation \citep[e.g.][]{Dekel2023}. The high electron densities result in collisional de-excitation of several commonly seen emission lines (i.e., [OII]) and lead to collisional excitation of several He I and H I lines. 

The strong UV lines provide additional insight into CNO abundances, revealing a strikingly different picture from typical star-forming galaxies at lower redshifts.
While the ionized ISM is characterized by an extremely low oxygen abundance ($[\mathrm{O/H}] = -1.3$) and the UV lines indicate a sub-solar C/O ($[\mathrm{C/O}] = -0.6$; consistent with the locus of star-forming galaxies at lower redshift at similar O/H; \citealt{Berg2019}), the prominence of nitrogen lines suggests a remarkably high N/O.
In particular, we infer from UV line ratios a value of $ \log(\rm N/O)=-0.39^{+0.10}_{-0.08}$, or $[\mathrm{N/O}] = 0.47$ --- an overabundance of N relative to O that is $3\times$ the solar value. 
This places \rxc{} in striking contrast with the plateau and increasing trend in N/O-O/H observed in local \ion{H}{ii} regions and typical star-forming galaxies --- but remarkably close to the values now inferred in several other luminous $z\gtrsim 6$ galaxies  \citep[Figure~\ref{fig:abundances};][]{Cameron2023,Senchyna2023,Isobe2023}.

There have been several other nitrogen-enhanced galaxies recently discovered by {\it JWST} in the reionization era. The most robust of these is GNz11 \citep{Bunker2023}, but tentative signatures are also seen in two other $z\gtrsim 6$ galaxies \citep{Isobe2023, Marques-Chaves2023, Larson2023}. 
As discussed above, the star formation conditions in RXCJ2248-ID appear broadly similar (and perhaps even more extreme) than GNz11, with both likely having recently experienced the onset of a strong burst of star formation and now harboring a dense concentration of massive stars. 
The spectral properties are also strikingly similar, with both showing a hard radiation field, high electron densities, and metal poor gas with strong nitrogen-enhancement. 
In GNz11, there is some ambiguity 
over the interpretation of the nitrogen features owing to potential signatures of AGN activity \citep{Maiolino2023}, including [Ne IV] emission, evidence for fast-moving outflows ($\gtrsim$ 1000 km/s), and indications of BLR-like densities ($10^{10}$ cm$^{-3}$) from [NIII] flux ratios. None of these potential AGN signatures is present in our spectrum of RXCJ2248-ID. 
While we cannot rule out some contribution from AGN photoionization in RXCJ2248-ID, the data do not indicate the same signatures of AGN activity as are seen in GNz11.  
Thus in the case of RXCJ2248-ID, the atypical spectral features (i.e., high electron density, hard radiation field, and nitrogen enhancement) appears more unambiguously connected to the dense population of recently formed massive stars.

As has been pointed out, the elevated N/O ratio inferred in \rxc{} and other luminous $z\gtrsim 6$ galaxies is not entirely anomalous in the broader context of the stellar archaeological record.
A defining feature of globular clusters is the appearance of abundance variations among their constituent stars, including a ubiquitous collection of strongly nitrogen-enhanced, oxygen-depleted stars which overlap the region of N/O-O/H space occupied by these integrated galaxy ISM measurements \citep[Figure~\ref{fig:abundances};][]{Senchyna2023,Charbonnel2023,Marques-Chaves2023,Isobe2023}.
While rarely encountered beyond globular clusters, similar enrichment patterns have been reported in a handful of other dense stellar environments, including ultracompact dwarfs and potentially the most ancient component of the Milky Way halo \citep{Belokurov2023}.
Such N/O and N/C enhancement is a characteristic signature of high-temperature nuclear burning of hydrogen via the CNO process; and other abundance ratios encoded in these cluster stars further evince their likely formation from gas polluted by the products of high-temperature nuclear burning \citep[e.g. see reviews by][]{Gratton2012,Bastian2018}.

While remarkably well-characterized across globular clusters, the precise origin of this nuclear-processed material and the mechanism by which it is ejected and allowed to subsequently cool and form stars remain hotly debated.
Essentially all proposed origins are naturally related to high-density clustered star formation; from massive star envelopes unbound by dynamically-enhanced binary interactions or winds, to longer-timescale AGB star ejecta captured and cooled by the deep potential well, to supermassive stars formed via a `conveyer belt' of collisions which could forge and eject the requisite material \citep[see][for a review]{Bastian2018}.
In \rxc{} we observe clear signatures of anomalously dense conditions of star formation, evident both from the compact spatial concentration of the light after lensing and signatures of strikingly high electron densities in the highly-ionized nebular gas phase.
While surprising, the detection of abundances reflective of CNO-processing in this gas alongside signs of high densities is also qualitatively consistent with the view that we are observing some stage of the extraordinarily dense conditions of an enriched second generation of stars forming under globular cluster-like conditions.

Many questions remain about the nature of \rxc{} and the growing list of other objects in which similar signatures of dense nitrogen-enriched gas has been reported.
One surprise is the apparent scale of the star formation events in both \rxc{} and \gnz{}.
While the SFR and stellar mass estimate of \gnz{} is particularly uncertain given the possible AGN contribution to the light, \rxc{} shows no clear evidence of non-stellar activity and yet also appears to host a burst of star formation of similarly high mass; $10^8$~$M_\odot$.
This is several orders of magnitude above the characteristic stellar mass scale of remnant globular clusters today.
This apparent discrepancy in mass scale could be interpreted as evidence that we are viewing a system of multiple globular clusters in formation together, potentially diluted by the light of other clusters.
This also might indicate we are observing a similar dense clustered star formation event but on a much larger scale, akin to that which may have taken place in the early Milky Way \citep[e.g.][]{Belokurov2023}.
However, it is also interesting to note that some lines of evidence in the study of some formation channels of globular clusters (in particular, those in which the material is produced by `normal' massive stars or AGB stars) have long predicted that their precursors may actually be orders of magnitude more massive than their remnants \citep[e.g.][]{Conroy2012,Cabrera-Ziri2015}.
Only limited speculation can be made on the basis of the small number of objects yet in-play.
But with larger demographic constraints and deeper spectroscopy from continued high-resolution rest-UV spectroscopic follow-up of lensed galaxies with \jwst{}, more direct ties between these signatures and the formation picture of globular clusters will become increasingly tractable.

%
%
%
%
\section{Summary}
\label{sec:summary}


Nearly ten years ago,  ground-based spectroscopy revealed the presence of very strong nebular CIV emission in what appeared to be typical low mass $z\gtrsim 6$ galaxies, suggesting hard radiation fields may be common in the reionization era. The ionizing sources and gas conditions linked to this hard radiation phase have long remained unclear owing to the limitations of ground-based spectroscopy at $z\gtrsim 6$. Here we present deep {\it JWST}/NIRSpec R=1000 spectroscopy and NIRCam imaging of the gravitationally-lensed galaxy RXCJ2248-ID, one of two $z\gtrsim 6$ CIV emitters known prior to {\it JWST}. 
We summarize our key findings below.

(i) We have presented deep {\it JWST}/NIRSpec $R\sim1000$ spectroscopy for three lensed images of the CIV emitter RXCJ2248-ID.
Each of the observed images is bright (J=24.8--25.9) and significantly magnified; the combination of individual spectra provide an extremely sensitive view of the system, with continuous wavelength coverage from the rest-UV through the rest-optical.
The combined spectrum serves as an exceptional template for nebular emission powered by low-metallicity massive stars in the reionization era.

(ii) 
RXCJ2248-ID is composed of two components separated by $\sim220$pc that are each unresolved ($r_{\rm e}\lesssim22 $ pc). Our spectral analysis focuses on the CIV-bright star forming complex. 
The SED of this clump demonstrates that the galaxy is in a period of extremely rapid assembly, with an SFR ($63^{+37}_{-23}$ M$_{\odot}$ yr$^{-1}$) that is among the highest found for $z\sim6-9$ galaxies with similar stellar masses ($1.1\times10^8$ M$\odot$). 
The high sSFR drives a population of massive stars within RXCJ2248-ID that is extremely compact, with an SFR surface density of $\Sigma_{\rm SFR}>10,400\rm ~M_{\odot}/yr/kpc^2$ and stellar mass surface density of $\Sigma_{\rm M_{*}}>3.6\times10^{10}\rm ~M_{\odot}/kpc^2$ (see Figure~\ref{fig:sizes}).
These densities are quite rare among the broader reionization era galaxy population, perhaps reflecting a short-lived phase. However we are detecting a growing number of systems (e.g., GN-z11) enhanced values.

(iii) 
The rest-frame UV spectrum of RXCJ2248-ID is dominated by high-ionization emission lines.
The CIV$\lambda1548$ ($\rm EW=34.1^{+0.8}_{-0.7}\angstrom{}$), HeII$\lambda1640$ ($\rm EW=5.6^{+0.4}_{-0.4}\angstrom{}$), and CIII]$\lambda1908$ ($\rm EW=21.7^{+0.7}_{-0.7}\angstrom{}$) are significantly stronger than what is typically observed among low-metallicity compact dwarf galaxies in the local Universe.
The ratios of rest-UV lines $\log(\rm CIV/HeII)=1.02^{+0.08}_{-0.10}$, $\log(\rm CIV/CIII])=0.44^{+0.05}_{-0.05}$, and $\log(\rm OIII]/HeII)=0.59^{+0.09}_{-0.10}$ constrain the shape of the ionizing spectrum, which is consistent with being of by stellar origin and does not require the presence of an AGN.
The absence of lines at yet higher ionization energies (e.g., [Ne IV]) further supports a stellar origin of the nebular spectrum.

(iv) The rest-optical spectrum of RXCJ2248-ID displays very large EWs ($\rm EW_{\rm [OIII]}=2800^{+115}_{-87}\angstrom{}$, $\rm EW_{\rm H\alpha}= 826^{+41}_{-43}\angstrom{}$) that indicate nebular-dominated emission powered by a very young stellar population. 
The rest-optical line ratios (O32=$184^{+88}_{-37}$, Ne3O2=$16.6^{+8.3}_{-3.3}$) reflect extremely intense ionization conditions, and are $\sim100$ and $\sim20$ times greater than typical systems at $z\sim2-3$ and in the reionization era, respectively.
RXCJ2248-ID displays very weak [OII] in addition to boosted emission of H I and He I lines which both likely are affected by collisional effects imparted due to high ionized gas densities.
We note that the impact of ISM conditions on key observables highlights the importance of high-quality templates such as RXCJ2248-ID to aid in interpreting reionization-era spectra.

(v) The ionized gas of RXCJ2248-ID is dominated by  high electron densities that range from $6.4-31\times10^4$ cm$^{-3}$, scaling  with the ionization energy of the density indicator.
The near absence of low ionization lines (e.g., [OII], [SII]) imply such regions minimally contribute to the nebular emission. We suggest the high ionized gas densities are a short phase reflecting the high gas densities required to power the strong burst of star formation.

(vi) The ionized gas is metal poor ($Z\sim1/20Z_{\odot}$), with an oxygen abundance of $12+\log(\rm O/H)=7.43^{+0.17}_{-0.09}$.  As is commonly seen in metal poor galaxies, the C/O 
ratio is inferred to be sub-solar. However the detection of strong NIV] ($\rm EW=17.3$\angstrom{}) and [NIII] ($\rm EW=4.3$\angstrom{}) suggests RXCJ2248-ID is significantly nitrogen-enriched, with an abundance that is $1$~dex above typical metal poor galaxies.  This level of nitrogen enhancement is consistent with expectations for CNO-processed gas. While this abundance pattern is almost never seen in lower redshift galaxies, it is seen in  globular cluster stars and also  mimics the abundance pattern recently identified in GNz11 \citep[e.g.,][]{Maiolino2023,Cameron2023, Senchyna2023}. Unlike GNz11, we do not detect clear AGN signatures in RXCJ2248-ID.

(vii) Our observations suggest that the nitrogen-enhancement and hard radiation field seen in some $z\gtrsim 6$ galaxies may be associated with a powerful burst of star formation that yields an extremely dense stellar environment.  The star formation conditions in RXCJ2248-ID appear broadly similar to that seen in GNz11. We suggest this may be a short-lived phase that becomes increasingly common in the reionization era. We discuss the potential connection of this phase to the formation of the 
second generation stars in globular clusters.

\section*{Acknowledgements}
This work is based in part on observations made with the NASA/ESA/CSA James Webb Space Telescope. The data were obtained from the Mikulski Archive for Space Telescopes at the Space Telescope Science Institute, which is operated by the Association of Universities for Research in Astronomy, Inc., under NASA contract NAS 5-03127 for JWST. These observations are associated with program 2478.

MWT acknowledges support from the NASA ADAP program through the grant number 80NSSC23K0467.
DPS acknowledges support from the National Science Foundation through the grant AST-2109066. 
AZ acknowledges support by Grant No. 2020750 from the United States-Israel Binational Science Foundation (BSF) and Grant No. 2109066 from the United States National Science Foundation (NSF); by the Ministry of Science \& Technology, Israel; and by the Israel Science Foundation Grant No. 864/23.

 \section*{Data Availability}
The data underlying this article may be presented upon reasonable request to the corresponding author.

\bibliographystyle{mnras}
\bibliography{main}

\end{document}